\documentclass[sort,compress]{els-mrw}

\usepackage{amsmath,amssymb,amsfonts,amsthm,makeidx,graphicx}
\usepackage{txfonts}
\usepackage{helvet}
\usepackage{slashed}
\usepackage{braket}
\usepackage{doi}
\usepackage{placeins}
\usepackage[normalem]{ulem}

\usepackage[font=small, labelfont=bf, textfont={small,it}]{caption}

\graphicspath{{FIGS/}}

\newcommand{\beq}{\begin{eqnarray}}
\newcommand{\eeq}{\end{eqnarray}}
\newcommand{\beqnn}{\begin{eqnarray*}}
\newcommand{\eeqnn}{\end{eqnarray*}}
\newcommand{\bal}{\begin{aligned}}
\newcommand{\eal}{\end{aligned}}

\newcommand{\Tr}{\mathrm{Tr}}
\newcommand{\CP}{\mathrm{CP}}
\newcommand{\SU}{\mathrm{SU}}
\newcommand{\U}{\mathrm{U}}

\newcommand{\dd}{\mathrm{d}}
\newcommand{\ii}{\mathrm{i}}
\newcommand{\ee}{\mathrm{e}}

\newcommand{\subYM}{{\scriptscriptstyle{\mathrm{YM}}}}
\newcommand{\subQCD}{{\scriptscriptstyle{\mathrm{QCD}}}}
\newcommand{\subquark}{{\scriptscriptstyle{\mathrm{q}}}}

\newcommand{\supYM}{{\scriptscriptstyle{(\mathrm{YM})}}}

\newcommand{\supquark}{{\scriptscriptstyle{(\mathrm{q})}}}

\newcommand{\supL}{{\scriptscriptstyle{(\mathrm{L})}}}

\newcommand{\pp}{{\scriptscriptstyle{(+)}}}
\newcommand{\mm}{{\scriptscriptstyle{(-)}}}

\renewcommand{\top}{{\scriptscriptstyle{\mathrm{top}}}}
\renewcommand{\S}{{\scriptscriptstyle{\mathrm{S}}}}
\newcommand{\eff}{{\scriptscriptstyle{\mathrm{eff}}}}
\newcommand{\subf}{{\scriptscriptstyle{\mathrm{f}}}}
\newcommand{\subA}{{\scriptscriptstyle{\mathrm{A}}}}
\newcommand{\subI}{{\scriptscriptstyle{\mathrm{I}}}}
\newcommand{\subL}{{\scriptscriptstyle{\mathrm{L}}}}
\newcommand{\subR}{{\scriptscriptstyle{\mathrm{R}}}}
\newcommand{\subE}{{\scriptscriptstyle{\mathrm{E}}}}
\newcommand{\supE}{{\scriptscriptstyle{(\mathrm{E})}}}
\newcommand{\subG}{{\scriptscriptstyle{\mathrm{G}}}}

\newcommand{\subN}{{\scriptscriptstyle{\mathrm{N}}}}
\newcommand{\subPQ}{{\scriptscriptstyle{\mathrm{PQ}}}}
\newcommand{\subeff}{{\scriptscriptstyle{\mathrm{eff}}}}

\newcommand{\EDM}{{\scriptscriptstyle{\mathrm{EDM}}}}
\newcommand{\EM}{{\scriptscriptstyle{\mathrm{EM}}}}

\begin{document}

\chapter{Topological Susceptibility and QCD at Finite Theta Angle}

\author[1]{Claudio Bonanno}
\address[1]{\orgname{Albert Einstein Center for Fundamental Physics, Institute for Theoretical Physics, University of Bern}, \orgaddress{Sidlerstra{\ss}e 5, CH-3012 Bern, Switzerland}}
%\ead{claudio.bonanno@unibe.ch}

\author[2,3]{Claudio Bonati}
\address[2]{\orgname{Dipartimento di Fisica, Universit\`a di Pisa}, \orgaddress{Largo Bruno Pontecorvo 3, I-56127 Pisa, Italy}}
\address[3]{\orgname{INFN - Sezione di Pisa}, \orgaddress{Largo Bruno Pontecorvo 3, I-56127 Pisa, Italy}}
%\ead{claudio.bonati@unipi.it}

\author[2,3]{Massimo D'Elia}
%\ead{massimo.delia@unipi.it}

\maketitle

\begin{abstract}[Abstract]
In this chapter we provide a pedagogical introduction to the main theoretical aspects related to topology and $\theta$-dependence in Quantum Chromo-Dynamics (QCD), and to their phenomenological relevance in the Standard Model ($\eta^\prime$ physics, neutron electric dipole moment) and beyond (strong CP problem and the axion solution). We then provide an overview of the main analytic predictions for $\theta$-dependence obtained using several different approaches (chiral effective theories, large-$N$ arguments, semiclassical methods) and their regimes of validity, as well as a selection of the most recent numerical results about QCD topology obtained via Monte Carlo simulations of the lattice-discretized theory.
\end{abstract}

\begin{keywords}
$\theta$-dependence \sep Strong CP problem \sep Topological susceptibility \sep Non-perturbative phenomena \sep Lattice QCD
\end{keywords}

%%%%%%%%%%%%%%%%%%%%%%%%%%%%%%%%%%%%%%%%%%%%%%%%%%%%%%%%%%%%%%%%
\begin{glossary}[Nomenclature]
\begin{tabular}{@{}lp{34pc}@{}}
$\chi$PT & Chiral Perturbation Theory\\
DIGA & Dilute Instanton Gas Approximation\\
I~/~A  & Instanton~/~Anti-instanton\\
PQ  & Peccei--Quinn\\
QFT & Quantum Field Theory\\
QCD & Quantum Chromo-Dynamics\\
WV & Witten--Veneziano \\
YM & Yang--Mills
\end{tabular}
\end{glossary}

\newpage

%%%%%%%%%%%%%%%%%%%%%%%%%%%%%%%%%%%%%%%%%%%%%%%%%%%%%%%%%%%%%%%%
%% the following item is mandatory: 
%% List of the key points and topics a reader can expect to learn from this chapter 
\section*{Objectives}
\begin{itemize}
\item In Sec.~\ref{sec:QCD_topo_intro} we introduce topology,
$\theta$-dependence, and the axion solution to the strong CP problem in a
pedagogical way.
\item In Sec.~\ref{sec:analytic_res} we summarize the main analytic predictions
about $\theta$-dependence based on large-$N$ considerations, on chiral
effective theories, and on semiclassical methods.
\item In Sec.~\ref{sec:num_res} we discuss $\theta$-dependence in lattice field theory, outlining the main numerical challenges involved, the strategies to address them, and reviewing recent results obtained in QCD and Yang–Mills theories.
\end{itemize}

%%%%%%%%%%%%%%%%%%%%%%%%%%%%%%%%%%%%%%%%%%%%%%%%%%%%%%%%%%%%%%%%

\section{Introduction}\label{sec:intro}

Quantum Chromo-Dynamics (QCD) is the theory describing the strong interactions among quarks and gluons in the Standard Model. One of the most fascinating aspects of this Quantum
Field Theory is the remarkable number of highly non-trivial features it
displays in its strongly-coupled low-energy regime. Among them one finds, for
example, color confinement, chiral-symmetry breaking, and $\theta$-dependence, which is the subject of this chapter.

The $\theta$ parameter is a genuine coupling of QCD, just like the strong
coupling constant, and from the mathematical point of view it is related to the topological features of the theory, as it couples to the winding number $Q$ of the gluon field: $\mathcal{S} \to \mathcal{S}(\theta)=\mathcal{S}+\theta Q$. The winding number $Q$ is a topological invariant related to the non-trivial features of the gauge group
manifold, and for appropriate boundary conditions it can only assume integer
values; at the same time, $Q$ is odd under P and CP, so that a non-zero
value of $\theta$ breaks these symmetries explicitly. The $\theta$-term
has a few other peculiar traits: since it can be written as the space-time
integral of a four-divergence, it does not affect the classical equations of
motion, and even in the quantum theory it has no effect at any order of
perturbation theory. Nevertheless, it has non-perturbative consequences in the quantum theory. This is best seen in the functional integral formalism, where the QCD path integral can be decomposed in a sum on the topological sectors of the gluon field (i.e., on the values of $Q$), with the $\theta$ parameter appearing as a complex phase weighting each sector differently, leading eventually to a non-trivial $\theta$-dependence of all observables.

QCD $\theta$-dependence is not only fundamental for its phenomenological
implications in the Standard Model, it is also crucial for several beyond the
Standard Model theories. The non-trivial $\theta$-dependence of the QCD vacuum (or thermal state, at finite temperature), parameterized at leading order in $\theta$ by the so-called \emph{topological susceptibility}, lies at the core of one of the most important open problems of the Standard Model, the \emph{strong CP problem}, and is key to a possible extension of the Standard Model proposed to solve it, the Peccei--Quinn axion. In this context, the study of QCD $\theta$-dependence plays a fundamental role, and has been the target of several theoretical studies. Since topology stems from purely non-perturbative gauge dynamics, non-perturbative tools must be used to reliably address $\theta$-dependence.

The goal of this chapter is essentially twofold. First, we aim at providing a
pedagogical introduction to the $\theta$ parameter in QCD and to its
phenomenological relevance (Sec.~\ref{sec:QCD_topo_intro}). Then, we aim at providing a compact summary of the main recent results about $\theta$-dependence in QCD and QCD-like theories, highlighting their theoretical and phenomenological significance (an extensive review of earlier results can instead be found in Ref.~\cite{Vicari:2008jw}). Analytic results obtained by using chiral effective theories, large-$N$ arguments and semiclassical methods will be discussed in Sec.~\ref{sec:analytic_res}, while those coming from numerical lattice simulations will be the subject of Sec.~\ref{sec:num_res}. Finally, we will present our conclusions by outlining key research directions currently being pursued by several groups (Sec.~\ref{sec:conclu}). At the end of the chapter, we also provide a summary table, Tab.~\ref{tab:summary_main_results}, where we collected the main numerical lattice QCD results discussed in Sec.~\ref{sec:num_res}.

\section{The \texorpdfstring{$\theta$}{theta} angle in QCD}\label{sec:QCD_topo_intro}

The Quantum Field Theory describing QCD is a $\SU(N)$ Yang--Mills theory, with
$N=3$ the number of colors, coupled to $N_{\subf}=6$ quark flavors transforming
in the fundamental representation of the color gauge group. Its Lagrangian in
Euclidean time reads:
\beq
\mathcal{L}_{\subE} = \frac{1}{2g^2} \Tr\left[G_{\mu\nu}(x)G_{\mu\nu}(x)\right] + \sum_{\rm f \, = \, 1}^{N_{\subf}} \overline{\psi}_{\subf}(\slashed{D}+m_{\subf})\psi_{\subf} \equiv \mathcal{L}_{\subYM} + \mathcal{L}_{\subquark} \ .
\eeq
Here $G_{\mu\nu}=\partial_\mu A_\nu - \partial_\nu A_\mu + \ii [A_\mu,A_\nu]$
is the gluon field strength, $A_\mu = A_\mu^a T_a$ with $\Tr[T_a T_b] =
\frac{1}{2}\delta_{ab}$ is the gauge field, $\psi_{\subf}$ is the quark field
with flavor f, $g$ is the strong coupling. Finally, $\slashed{D} =
\gamma^{\supE}_\mu D_\mu$, with $D_\mu = \partial_\mu + \ii A_\mu$ the
covariant derivative and $\gamma_\mu^{\supE}$ the Euclidean Dirac matrices.
Let us start our discussion on the topological features of QCD by introducing
the topological charge density $q(x)$, and its space-time integral, $Q$:
\beq\label{eq:gluon_charge_def}
q(x) = \frac{1}{16\pi^2} \Tr\left[G_{\mu\nu}(x)\widetilde{G}_{\mu\nu}(x)\right], \qquad Q = \int q(x) \,\dd^4 x \, \, .
\eeq
Here, $\widetilde{G}_{\mu\nu}(x) \equiv
\frac{1}{2}\varepsilon_{\mu\nu\rho\sigma}G_{\rho\sigma}(x)$ is the dual field
strength tensor, and we recall that one
can write the gauge-invariant quantity $q(x)$ as the four-divergence of
the gauge-non-invariant Chern--Simons current $K_\mu(x)$:
\beq
q(x) = \partial_\mu K_\mu(x) \,; \ \ \ \ \ K_\mu \equiv
\frac{1}{16 \pi^2} \varepsilon_{\mu\nu\rho\sigma} A_\nu^a
\left( \partial_\rho A_\sigma^a + \frac{1}{3} f^{abc} A_\rho^b A_\sigma^c \right) \,; \ \ \ \ \ [T_a,T_b] = \sum_c f^{abc} \,T_c \,,
\eeq
with $f^{abc}$ the $\SU(N)$ structure constants.

Semiclassically, and choosing suitable boundary conditions, $Q$ can be shown to
be equal to the number of times the gauge field covers the gauge group at
infinite distance\footnote{A theoretically sound interpretation beyond
semiclassics can also be given in finite volume within the Hamiltonian
formalism, see, e.g., the recent pedagogical review~\cite{Bonanno:2025wcv}.};
in such conditions, the space of gauge fields with finite Euclidean action
splits into topological sectors, each one identified by a different value of
$Q$.  The best known construction is obtained imposing
$G_{\mu\nu}(x\to\infty)=0$ as a boundary condition. In this case, the gauge
field approaches a gauge copy of zero at infinite distance,
$A_\mu(x)\underset{r\to\infty}{\to}\Omega(\hat{r})\partial_\mu
\Omega^\dagger(\hat{r})$, where $\hat{r}=x/r$ denotes the direction along which
infinity is approached.
Considering $\SU(2)$ as the paradigmatic case, one immediately sees the
role played by the non-triviality of the third homotopy group, $\pi_3$, of the
gauge group. On the one hand, $\hat{r}$ can be mapped onto a point on the
3-sphere S$^3$; on the other hand, $\Omega\in\SU(2)\simeq\mathrm{S}^3$, as it
is manifest from the well-known parameterization:
\beq
\Omega = a_0 \mathrm{I} + \vec{a} \cdot\vec{\sigma}, \qquad a_0^2 + \vert\vec{a}\vert^2 = 1, \qquad \mathrm{I} = \text{ identity matrix}, \quad \vec{\sigma} = \text{ Pauli matrices} \, .
\eeq
Thus, $\Omega(\hat{r})$ is a map from the 3-sphere onto itself. Maps $\mathrm{S}^3\to\mathrm{S}^3$ which can be continuously deformed
into each other constitute an equivalence class (homotopy class), and are
labeled by the integer winding number counting how many times $\Omega$ wraps
around the 3-sphere in color space when the 3-sphere in coordinate space is
covered once. Mathematically, what we have said
so far is succinctly expressed as $\pi_3(\mathrm{S}^3)=\mathbb{Z}$. The
topological charge $Q$ in Eq.~\eqref{eq:gluon_charge_def} can be shown to be
exactly an integral representation of this winding number.

A useful way to show the relevance of the topology in characterizing the QCD vacuum is offered by the Bogomolny inequalities:
\beq
\frac{1}{2g^2}\int \dd^4 x \, \Tr\left[G_{\mu\nu}(x) G_{\mu\nu}(x)\right]
= \frac{1}{4 g^2}\int \dd^4 x \, \Tr\left[(G_{\mu\nu}(x) \pm \widetilde G_{\mu\nu}(x) ) (G_{\mu\nu}(x) \pm \widetilde G_{\mu\nu}(x))
\right] \mp \frac{1}{2g^2}\int \dd^4 x \, \Tr\left[G_{\mu\nu}(x) \widetilde G_{\mu\nu}(x) \right]
\geq \frac{8 \pi^2}{g^2} \vert Q \vert \, , 
\eeq
In each sector these bounds are saturated by gauge fields
satisfying $\widetilde G_{\mu\nu} = \pm\, G_{\mu\nu}$, the so-called self-dual
(+) and anti-self-dual (-) conditions. Since self-dual/anti-self-dual
fields are local minima of the Euclidean action for
the gluon field, with $\mathcal{S}_{\min} = 8 \pi^2 \vert Q \vert/g^2$,
they are solutions of the Yang--Mills classical equations of motion. Solutions to $\widetilde
G_{\mu\nu} = \pm\, G_{\mu\nu}$ have been analytically derived in
Refs.~\cite{Belavin:1975aaa,Achtor:1979aaa}, and for $Q=\pm 1$ they are usually called
instantons/anti-instantons. The expression for the SU(2) instanton is given by:
\beq
A_{\mu}(x) = \eta^{a}_{\mu\nu} \frac{(x-c)_\nu}{\vert x - c\vert^2 + \rho^2}\sigma_a, \qquad\qquad \eta^a_{\mu\nu} = \varepsilon_{0a\mu\nu} + \delta_{a\mu}\delta_{\nu 0} - \delta_{a \nu}\delta_{\mu 0},  \qquad \qquad \text{($\SU(2)$ instanton $Q=1$ )},
\eeq
where $c$ is the instanton position, $\rho$ is the instanton size, and
$\{\sigma_a\}_{a\,=\,1}^{3}$ are the Pauli matrices. The fact that
$G_{\mu\nu}(x)$ is significantly different from zero only in a neighborhood of
typical linear size $\rho$ of $c_{\mu}$ (i.e., a short amount of time in
a specific position) justifies the ``instanton'' name and the interpretation of the instanton field as describing a pseudo-particle, see Sec.~\ref{subsect:DIGA}.
By substituting $\eta^a_{\mu\nu} \longrightarrow \overline{\eta}^{\,
a}_{\mu\nu}= \varepsilon_{0a\mu\nu} - \delta_{a\mu}\delta_{\nu 0} + \delta_{a
\nu}\delta_{\mu 0}$ one instead obtains the $Q=-1$ $\SU(2)$ anti-instanton.
Generalizations to $N>2$ can be obtained embedding $N=2$ solutions into the $\SU(2)$ subgroups of $\SU(N)$, thus the explicit form of the instanton solution also depends on parameters related to its color orientation.

Being the topological charge density a dimension 4 gauge-invariant and
Lorentz-invariant operator, a term proportional to $q(x)$ can be added to the 
standard QCD Lagrangian without spoiling the renormalizability of the theory
:
\beq \label{eq:thetaterm}
\mathcal{L}_{\subE}(\theta) = \mathcal{L}_{\subE} - \ii \theta q \, .
\eeq
This so-called $\theta$-term introduces a new coupling of QCD, the
$\theta$ parameter, and it is purely imaginary in the Euclidean formulation,
due to the odd number of time derivatives appearing in the product $G
\widetilde G$.  Despite not contributing to the classical equation of motion,
since $q$ is a total derivative, the topological term nonetheless
introduces a non-trivial dependence on $\theta$ in the quantum theory, with
deep theoretical and phenomenological consequences. A convenient way of showing
this is to consider the finite temperature partition function of QCD, $Z(T,\theta)$, in
the functional integral formalism:
\beq
Z(T,\theta) = \hspace{-2pt} \int [\dd A \, \dd \overline{\psi} \, \dd \psi] \, \ee^{-\mathcal{S}_{\subE}+\ii\theta Q}
= \hspace{-2pt} \int [\dd A] \prod_{\mathrm{f}\,=\,1}^{N_{\subf}}\det\left[\slashed{D}[A]+m_{\subf}\right] \, \ee^{-\mathcal{S}_{\subE}^{\supYM} + \ii \theta Q} = \hspace{-3pt} \sum_{n\,=\,-\infty}^{\infty} 
\ee^{\ii n \theta} Z_n(T) \, , \ \ \ \ \ \ \ \ \ \ \mathcal{S}_{\subE} = \hspace{-2pt} \int_0^{1/T} \hspace{-5pt}
\dd \tau \int \hspace{-2pt} \dd^3 x \, \mathcal{L}_{\subE} = \mathcal{S}_{\subE}^{\supYM} + \mathcal{S}_{\subE}^{\supquark}\, .
\label{eq:ZthetaQCD}
\eeq
Here, $\mathcal{S}_{\subE}$ is the Euclidean QCD Lagrangian defined on a
space-time manifold with a compactified temporal direction of length $1/T$, and
periodic/anti-periodic temporal boundary conditions for gluon/quark fields. As
it can be seen, the partition function splits into the sum of an infinite
number of functional integrals restricted to a specific topological sector:
\beq
Z_n(T) = \int[\dd A] \, \prod_{\mathrm{f}\,=\,1}^{N_{\subf}}\det\left[\slashed{D}[A]+m_{\subf}\right] \ee^{-\mathcal{S}^{\supYM}_{\subE}} \, \delta(Q[A]-n) \, ,
\eeq
and the $\theta$-term contribution manifests as a complex phase weighting
differently each $Z_n(T)$. At this stage, it is also worth commenting that the
instanton path-integral weight is $\sim \ee^{-1/g^2}$, thus purely
non-perturbative in the coupling $g$, thus pointing out the inherent
non-perturbative nature of the topological features of gauge theories.

A quantity tightly related to $Z(T,\theta)$, which will be central for the
following discussion, is the free energy density $f(T,\theta)$, and its
$\theta$-dependent part $F(T,\theta)$, defined as:
\beq
F(T,\theta) = f(T,\theta)-f(T,0) = -\frac{1}{V}\log\left[\frac{Z(T,\theta)}{Z(T,0)}\right]\, , \qquad V = \frac{T}{L^3} = \text{ four-dimensional volume}\, .
\eeq
Since $Q$ is odd under parity due to presence of
$\varepsilon_{\mu\nu\rho\sigma}$, and since parity is a symmetry of
$\mathcal{S}_{\subE}$ at $\theta=0$, we have
$F(T,\theta)=F(T,-\theta)$, i.e., the free energy density is
an even function of $\theta$. Moreover, since $\theta$ only enters $Z$ through the
phases $\ee^{\ii n \theta}$, $F(T,\theta)$ is $2\pi$-periodic, and, due to the
Vafa--Witten theorem, it possesses a global minimum in $\theta = 0$, which is
thus also expected to be an analytic point~\cite{Vafa:1984xg} (i.e., parity cannot
be spontaneously broken at $\theta=0$). Based on these considerations, it is customary to parameterize the $\theta$-dependent part of the free energy as follows~\cite{Vicari:2008jw}:
\beq
\label{eq:thetadep_free_energy_general}
F(T,\theta) = \frac{1}{2} \, \chi(T) \, \theta^2 \left(1+\sum_{n\,=\,1}^{\infty}b_{2n}(T)\theta^{2n}\right) \, .
\eeq
The leading order $\mathcal{O}(\theta^2)$ coefficient $\chi$ is known as the {\em topological susceptibility}:
\beq
\chi = \frac{\braket{Q^2}}{V}\Bigg\vert_{\theta \, = \, 0}, 
\eeq
and it is proportional to the variance of the
topological charge distribution at $\theta=0$, $\mathcal{P}(Q)$. Higher-order
coefficients in $\theta$ parameterize instead the deviations of
$\mathcal{P}(Q)$ from a pure Gaussian, and are related to higher-order cumulants
of $\mathcal{P}(Q)$, denoted as $\braket{Q^k}_c$:
\beq\label{eq:b2n_def}
b_{2n} = \frac{2(-1)^{n}}{(2n+2)!} \frac{\braket{Q^{2n}}_{c}}{\braket{Q^2}}\bigg\vert_{\theta\,=\,0} \, .
\eeq
The explicit expression of the first few cumulants is: 
\beq
\braket{Q^4}_{\rm c}\big\vert_{\theta\,=\,0} = \braket{Q^4} - 3\braket{Q^2}^2 \, , \ \ \ \ \ 
\braket{Q^6}_{\rm c}\big\vert_{\theta\,=\,0} &=& \braket{Q^6} - 15\braket{Q^4}\braket{Q^2} + 30 \braket{Q^2}^3 \, ,
\ \ \ \ \ \ \, \dots 
\eeq
where we have used that expectation values of all odd power of $Q$ identically
vanish at $\theta=0$, due to the parity symmetry.

Although so far we have introduced topology as a formal property of the gluon
field, it is actually a feature that is deeply connected with the chirality of
quarks via the well-known Atiyah--Singer index theorem:
\beq\label{eq:index_theorem_def}
Q[A] = n_0^{\scriptscriptstyle{(\rm L)}} - n_0^{\scriptscriptstyle{(\rm R)}},
\eeq
This equation expresses the fact that the massless Dirac operator, in the
background of a gauge field with a topological charge $Q[A]$, possesses a
number $n_0^{\scriptscriptstyle{(\rm L)}}$/$n_0^{\scriptscriptstyle{(\rm R)}}$
of chiral left-handed/right-handed zero eigenvalues, whose number is constrained by $Q$ via Eq.~\eqref{eq:index_theorem_def}. The relation between topology and chiral symmetry has fundamental implications for $\theta$-dependence. 
In the limit of $n_{\subf}$
massless quarks, the classical QCD Lagrangian is symmetric under the chiral
flavor group $\SU(n_{\subf})_{\scriptscriptstyle{\rm L}} \times
\SU(n_{\subf})_{\scriptscriptstyle{\rm R}} \times \U(1)_{\scriptscriptstyle{\rm
V}} \times \U(1)_{\scriptscriptstyle{\rm A}}$. Let us now focus on the Abelian
axial sub-group. It can be shown that, at the quantum level, this symmetry
becomes \emph{anomalous}~\cite{Adler:1969gk, Bell:1969ts, tHooft:1976rip}, as can be inferred, for
example, from the fact that the fermion measure in the functional integral is not invariant under a singlet axial transformation:
\beq
\begin{cases}
\psi_{\subL} &\underset{\U(1)_{\subA}}{\longrightarrow} \ee^{+\ii\alpha} \psi_{\subL}\\
\psi_{\subR} &\underset{\U(1)_{\subA}}{\longrightarrow} \ee^{-\ii\alpha} \psi_{\subR}
\end{cases}
\implies [\dd\overline{\psi}\,\dd \psi] \underset{\U(1)_{\subA}}{\longrightarrow} [\dd\overline{\psi}\,\dd \psi] \, \ee^{-\mathcal{A}}\, , \qquad \mathcal{A} = \ii 2 N_{\subf} \alpha Q\, , \qquad \psi_{\subL,\subR} = \frac{1\pm\gamma_5}{2} \psi\, , \qquad \psi=(\psi_1, \psi_2, \dots, \psi_{n_{\subf}}) \, .
\eeq
The quantity $\mathcal{A}$ is the \emph{chiral anomaly}, and to
write it in the previous form we used the index theorem. When quark masses are
non-vanishing, a singlet chiral transformation affects the action
$\mathcal{S}_{\subE}(\theta)$ as follows:
\beq
\label{eq:U1axial_transform}
\theta \underset{\U(1)_{\subA}}{\longrightarrow} \theta - 2 N_{\subf} \, \alpha\, ,\qquad
\mathcal{M} \underset{\U(1)_{\subA}}{\longrightarrow} \ee^{-2\ii\alpha} \mathcal{M}\, .
\eeq
Now, also the quark mass matrix
$\mathcal{M}=\mathrm{diag}(m_1,\dots,m_{N_{\subf}})$ changes under a
U(1)$_{\subA}$ rotation. Choosing $\alpha=\theta/(2N_{\subf})$ it is thus
possible to trade the $\theta$-term for a complex phase
$\ee^{-\ii\theta/N_{\subf}}$ in $\mathcal{M}$. However, it is not possible to
remove completely $\theta$-dependence from the theory since
\beq
\Theta \equiv \theta-\arg\det \mathcal{M}
\eeq
is invariant under the transformation in Eq.~\eqref{eq:U1axial_transform}. Removing the $\theta$-parameter via a $\U(1)_{\subA}$ field redefinition would be possible only if there was at least one massless flavor, as in that case there would be no change in the quark mass
term.

\subsection{The \texorpdfstring{$\U(1)_{\subA}$}{U(1)A} puzzle and the role of the topological susceptibility}\label{sec:WittenVeneziano}

As earlier outlined, a QCD-like theory with $n_{\subf}$ massless quarks
 is classically invariant under
$\SU(n_{\subf})_{\scriptscriptstyle{\rm L}} \times
\SU(n_{\subf})_{\scriptscriptstyle{\rm R}} \times \U(1)_{\scriptscriptstyle{\rm
V}} \times \U(1)_{\scriptscriptstyle{\rm A}}$ transformations. We have discussed that
$\U(1)_{\scriptscriptstyle{\rm A}}$ is anomalous. What is the fate of the other
sub-groups in the quantum theory? $\U(1)_{\scriptscriptstyle{\rm V}}$ 
is realized \emph{\'a la} Wigner at the quantum level,
and it is associated with the baryon number conservation. The
$\SU(n_{\subf})_{\scriptscriptstyle{\rm L}} \times
\SU(n_{\subf})_{\scriptscriptstyle{\rm R}}$ sub-group is instead spontaneously
broken to $\SU(n_{\subf})_{\scriptscriptstyle{\rm V}}$. Assuming
$n_{\subf}=3$, this implies the existence of $n_{\subf}^2 -1 = 8$ massless
Goldstone bosons. In the real world, there is no massless quark, but there are
three quarks (up, down and strange) whose mass is smaller than
$\Lambda_{\subQCD}$. Thus, their masses softly break the chiral-symmetry,
 and the would-be massless Goldstone bosons acquire a
light mass. These states are identified with the three pions $\pi^{\pm},\pi_0$, the
four kaons $K^0,\overline{K}^0,K^{\pm}$ and the $\eta$ meson.

The flavor-singlet $\eta^\prime$, instead, has a much larger mass then the components of the lightest octet. The explanation of this fact poses a conundrum since
S.~Weinberg pointed out that $m_{\eta^\prime} \le \sqrt{3}m_\pi\simeq 240$ MeV
were the origin of the $\eta^\prime$ mass related to a spontaneous chiral
symmetry breaking mechanism~\cite{Weinberg:1975ui}. The bound is largely
violated by the physical $\eta^\prime$ mass, and, although it was soon realized
that the violation was related to the anomalous breaking of the axial U(1)
symmetry by the anomaly~\cite{tHooft:1976rip}, the precise mechanism connecting these two aspects was not clear. For this reason, this issue was known as the $\U(1)_{\subA}$ puzzle. The resolution of this puzzle was provided by Witten and Veneziano and it heavily relies on the non-trivial $\theta$-dependence of QCD.

The Witten--Veneziano solution to the $\U(1)_{\subA}$ puzzle is based
on an ideal limit of QCD in which, at fixed number of flavors, the number of
colors is sent to infinity $N\to\infty$ and, at the same time, the strong
coupling constant $g^2$ is weakened, so that $g^2 N = \lambda$
remains constant as $N\to\infty$. In this limit, the QCD Euclidean action
assumes the peculiar form
\beq
S_{\subE} = \frac{N}{\lambda}\left\{ \frac{1}{4} \int \dd^4 x \, G_{\mu\nu}^a G_{\mu\nu}^a + \frac{\lambda}{N} \int \dd^4 x \, \overline{\psi}(\slashed{D}+\mathcal{M})\psi\right\}
\eeq
which suggests the possibility, at least in principle, of developing the functional integral as an asymptotic series in $1/N$ from a saddle-point approximation. In practice, carrying on such saddle-point expansion analytically in QCD is impossible; nonetheless, the $1/N$ expansion remains a very important framework to study the non-perturbative regime of strong interactions, as we shall see in this section. Let us also recall that the requirements $\lambda\sim\mathcal{O}(N^0)$ and $N_{\subf}\sim\mathcal{O}(N^0)$ are also
important to ensure that $\Lambda_{\subQCD}\sim\mathcal{O}(N^0)$, as one can
easily realize from the one-loop running of $g^2$:
\beq
\frac{g^2(\mu)}{4\pi} = \frac{4\pi}{\beta_0} \frac{1}{\log\left(\mu^2/\Lambda_{\subQCD}^2\right)}, \qquad \beta_0 = \frac{11}{3}N - \frac{2}{3}N_{\subf},
\eeq
i.e., in the large-$N$ limit with $N_{\subf}$ fixed, $\Lambda_{\subQCD}$ is
just a function of $g^2 N = \lambda$.

The $1/N$ expansion of QCD was
systematically analyzed in~\cite{tHooft:1973alw}, where it was shown that, in
the large-$N$ limit, quark contributions are suppressed by a factor of $1/N$
with respect to gluonic ones (a fact that could have also been naively guessed
by counting the color degrees of freedom for $A_\mu$ and $\psi$).  This is
however at odds with another fact earlier explained: the disappearance of
$\theta$-dependence in the chiral limit by virtue of the chiral anomaly.
Indeed, the anomaly is a purely fermionic contribution that thus should be
suppressed at large-$N$: yet, the disappearance of $\theta$-dependence
in the $m\to 0$ limit is a property that
holds for any value of $N$.  Witten and Veneziano solved this issue by
hypothesizing that there is one meson state (identified with the $\eta^\prime$)
whose squared mass scales as $1/N$, so that its propagator can contribute at
leading order at large $N$. This assumption allows to recover the vanishing of
the QCD topological susceptibility (and thus the absence of
$\theta$-dependence) by imposing a cancellation between the gluons' and the
$\eta^\prime$'s contributions to $\chi$ at leading order of the $1/N$
expansion. This eventually leads to the following well-known
formula~\cite{Veneziano:1979ec, Witten:1980sp, DiVecchia:1980yfw}:
\beq
m_{\eta^\prime}^2 = \frac{2 N_\subf}{F_\pi^2}\chi_{_\subYM}^\infty + 2 m_K^2 - m_{\eta}^2 \, .
\label{eq:wv_inverted}
\eeq
Here, $\chi_{_\subYM}^\infty$ represents the large-$N$ limit of the topological
susceptibility of the pure-glue theory. Unlike the $2 m_K^2 - m_{\eta}^2$
contribution, it remains finite in the chiral limit, thus leaving the
$\eta^\prime$ heavy in the limit in which the members of the lightest octet
become massless Goldstones. This clarifies how the non-trivial gluon topology
gives mass to the $\eta^\prime$ via the anomalous breaking of the axial
symmetry, and solves the $\U(1)_{\subA}$ puzzle. For this mechanism to work,
one must assume $\chi_{_\subYM}^{\infty}$ to be finite in the limit
$N\to\infty$. Exposing it in the previous equation, and using the experimental
values for the meson masses and the decay constant $F_\pi$, one finds (for
$N_{\subf}=3)$:
\beq
\chi_{_\subYM}^\infty = \frac{F_\pi^2}{2 N_\subf} \left( m_{\eta^\prime}^2 + m_{\eta}^2 - 2 m_K^2 \right) \approx \left(180\text{ MeV}\right)^4.
\label{eq:wv}
\eeq
This was the first case in which the $\theta$-dependence of QCD was clearly related to an important open problem in hadron phenomenology.

Another intriguing topological quantity, related to the Witten--Veneziano
solution of the $\U(1)_{\subA}$ puzzle, is the so-called topological
susceptibility slope. The starting point is the topological charge density
correlator in momentum space and its momentum expansion:
\beq
\widetilde{C}(p^2) = \int \dd^4 x \, \ee^{\ii p_\mu x_\mu} \, \braket{q(x)q(0)} = \chi - \chi^\prime p^2+\mathcal{O}(p^4)\, , \qquad p^2=p_\mu p_\mu\, .
\eeq
The leading order term at zero momentum is simply the topological
susceptibility by virtue of the translation invariance of the 2-point function
$C(x-y)=\braket{q(x)q(y)}$:
\beq
\int \dd^4 x \braket{q(x)q(0)} = \lim_{V\to\infty} \frac{1}{V} \int \dd^4 x \, \dd^4 y \, C(x-y) = \lim_{V\to\infty}\frac{\braket{Q^2}}{V}.
\eeq
The next-to-leading-order term is instead the so-called \emph{topological susceptibility slope}:
\beq
\chi^\prime = \frac{1}{8} \int \dd^4 x \vert x \vert^2 \braket{q(x)q(0)},
\eeq
which is proportional to the second moment of $C(x)$. 
In the derivation of the Witten-Veneziano formula we assumed that $\chi$ dominates the
momentum expansion of $\widetilde{C}(p^2)$ up to momenta $p^2 \sim m_{\eta^\prime}^2$. This is only true provided that:
\beq
\vert \chi^\prime \vert \ll \chi/m_{\eta^\prime}^2.
\eeq
This is thus yet another Yang--Mills topological quantity with a key connection to hadron phenomenology.

\subsection{The strong CP problem and the axion solution}

%The strong CP problem

%modello di bassa energia

%parametri del modello da cose note

%cose da fissare, connessione con DM, interesse per alta T, cose note e problemi

The topological charge is odd under a parity P transformation and even under a
charge conjugation C transformation, and thus odd under a combined CP
transformation. The CP-odd nature of the topological charge implies that the
$\theta$-term provides a source of strong CP violations. In particular, CP-odd
observables will acquire a non-vanishing vacuum expectation value at non-zero
$\theta$, a fact that can be used to estimate the value of $\theta$ (more
precisely, of $\Theta$) from experimental measurements of CP-odd hadronic
quantities. The most promising avenue to the experimental detection of strong CP violations is the measure of the neutron electric dipole moment (EDM), a CP-odd quantity that can
be measured experimentally with extremely high accuracy. At leading order
in $\theta$, dimensional analysis suggests the following estimate for
the neutron (N) EDM:
\beq\label{eq:NEDM_dimensional}
\vert d_{\subN} \vert \approx \vert \theta \vert \, e \, \frac{m_\pi^2}{m^3_{\subN}} \simeq 3.8 \cdot 10^{-3} \, \vert \theta\vert \, e \, \mathrm{fm} \, .
\eeq
This result has been further refined via effective theories and model
calculations~\cite{Pich:1991fq,Borasoy:2000pq,Hockings:2005cn,Narison:2008jp,Ottnad:2009jw,deVries:2010ah,Mereghetti:2010kp,Guo:2012vf,Bartolini:2016dbk},
with the most precise estimate coming from Chiral Perturbation
Theory~\cite{Guo:2012vf}:
\beq
d_{\subN} = (-2.9\pm 0.9) \cdot 10^{-3} \, \theta \, e \, \mathrm{fm} \, .
\eeq
So far, no signal above zero has been experimentally found for the neutron EDM,
with the most precise bound coming from the experiment conducted at the
PSI~\cite{Abel:2020pzs}:
\beq
\vert d_{\subN} \vert = (0.0\pm1.1_{\rm stat} \pm 0.2_{\rm syst})\cdot 10^{-13} \, e \, \mathrm{fm}.
\eeq
This in turn implies the tiny upper bound $\vert \theta\vert_{\rm exp} \lesssim
10^{-9}$ -- $10^{-10}$, which seems to suggest that the QCD $\theta$-parameter,
in order to describe experimental evidence, should be fine-tuned to the special
CP-conserving value $\theta=0$.  Generally speaking, this is problematic, as we
have seen that the chiral anomaly allows to reshuffle the gluonic topological
term into the quark mass matrix, thus opening a connection between CP
violations in the strong and in the electroweak sector: since such violations
are known to take place in the latter, what prevents their manifestation in the
former?  The only possible solution~\footnote{A few unconventional alternative solutions have been proposed in recent years. Some of these appear to present significant challenges~\cite{Ai:2020ptm,Ai:2024cnp,Nakamura:2021meh,Schierholz:2023hkx,Schierholz:2024var}, as evidenced by the vigorous debate they have generated within the community~\cite{Albandea:2024fui,Kaplan:2025bgy,Benabou:2025viy,Khoze:2025auv,Gamboa:2025hxa,Bhattacharya:2025qsk,Williams:2026cec,Aghaie:2026pkf,Ringwald:2026apz,Sannino:2026wgx}, others remain under active theoretical investigation~\cite{Strocchi:2024tis, Kaplan:2023pxd,Kaplan:2023pvd}.} to remove $\theta$-dependence within the Standard Model, as earlier anticipated, would be to have an exactly massless quark. This
scenario is sometimes called the ``massless up quark
solution''~\cite{Kaplan:1986ru,Choi:1988sy,Banks:1994yg}, due to the fact the
up quark is the lightest one. However, this solution has
been ruled out by Lattice
QCD~\cite{Alexandrou:2020bkd,FlavourLatticeAveragingGroupFLAG:2021npn}. The
difficulty of explaining why the QCD $\theta$ angle is equal to zero is
referred to as the \emph{strong CP problem}. This is one of the most important
open problems of the Standard Model, and it seems natural to look for a
solution of this issue beyond this framework.

In this respect, an attractive avenue is provided by the Peccei--Quinn
axion~\cite{Peccei:1977hh,Peccei:1977ur, Wilczek:1977pj, Weinberg:1977ma}. This
proposal relies on very few ingredients. One assumes the presence of an extra
axial global symmetry U(1)$_{\subPQ}$, spontaneously broken below a very high
scale $f_a$. At low energies, its associated pseudo Goldstone boson (the axion)
must couple to gluons via the topological charge density:
\beq
\mathcal{L}_{\subQCD} + \mathcal{L}_a \supset \left(\frac{a(x)}{f_a}+\theta\right)q(x) = \theta_{\subeff}(x)\, q(x)\, .
\eeq
This peculiar Lagrangian term is dictated by the axial nature of
U(1)$_{\subPQ}$, which makes it anomalous: the
axion-gluon coupling $\frac{a(x)}{f_a} q(x)$ is necessary to correctly
reproduce the axial anomaly under the action of the shift symmetry $a(x)/f_a
\to a(x)/f_a + \alpha$ the axion field enjoys by virtue of its Goldstone-boson
nature. All the other interaction terms of the axion with Standard Model
particles are in general not universal, however, they will all be suppressed by
powers of $f_a$ and involve derivatives of the
axion field. This means that, after integrating out
QCD, the temperature-dependent axion effective potential is given by:
\beq
\mathcal{V}_{\subeff}(T,a) = F(T,\theta_{\subeff}) \, ,
\eeq
and that the absolute minimum of $\mathcal{V}_{\subeff}(T,a)$,
towards which the axion field is driven, is reached for $\theta_{\subeff}=0$,
thus dynamically solving the strong CP problem.

Since the axion decay constant $f_a$ is expected to be very large, the axion
will have very feeble interactions with Standard Model particles. For this
reason, it is also regarded as a very promising Dark Matter candidate.
In this context, the study of the $\theta$-dependence of the QCD free energy,
and how it changes as a function of the temperature, is a crucial input to fix
much of axion physical properties. As an example, by matching
$\mathcal{V}_{\subeff}$ and $F$ order by order, one finds:
\beq
m_a^2 = \frac{\chi}{f_a^2}\, , \qquad \lambda_a = 12\, b_2 \frac{m_a^2}{f_a^2}\, ,
\eeq
with $m_a$ the axion mass, and $\frac{\lambda_a}{4!}a^4$ the axion quartic self-interaction.
Therefore, QCD results for $F(T,\theta)$ are necessary inputs to study the
cosmological evolution of axion Dark Matter, and in particular to estimate its relic
abundance today from the so-called \emph{misalignement mechanism} (see instead Sec.~\ref{sec:intro_sphaleron} for the axion \emph{thermal} production). In this
framework, one computes the axion evolution starting from an initial condition
where $\theta = \theta_0 \ne 0$ (assuming the Peccei--Quinn symmetry to be
restored at very high temperatures, few instants after the Big Bang). As the
Universe evolves and cools down, the temperature of the cosmological medium
drops, and the change of $\theta$-dependence of QCD as a function of the
tempearture induces an equivalent change in the shape of the axion potential.
This drives the axion field towards its minimum, realizing $\theta=0$ and
suppressing strong CP violations. This process is responsible for the
production of relic axions that could make (part of) the observed Dark
Matter~\cite{Preskill:1982cy,
Abbott:1982af,Dine:1982ah,Wantz:2009it,Berkowitz:2015aua} (see also the recent
review Ref.~\cite{DiLuzio:2020wdo}). In suitable temperature regimes, the
$\theta$-dependence of the free energy can be obtained analytically. However,
one has in general to resort to non-perturbative numerical simulations of
lattice QCD to be able to compute $F(T,\theta)$ from first principles. All
available results obtained with these analytical and numerical approaches will
be discussed in the next sections.

\subsection{Theoretical and phenomenological relevance of real-time QCD topological transitions}\label{sec:intro_sphaleron}

We have previously shown that four dimensional gauge configurations with
finite Euclidean action can be characterized by their topological charge $Q$,
and that for each value of $Q$ there are configurations (instantons) which minimize the
Euclidean action. These actions are  those which contribute the most to the
Euclidean path-integral in the semiclassical limit, however we have not yet discussed their physical interpretation. For this purpose it is convenient to fix 
the temporal gauge $A_0=0$. In this gauge, time independent vacuum states
can be characterized by the value of their \emph{three-dimensional} winding number $n$,
and one obtains a semiclassical picture of the Yang--Mills vacuum characterized
by several different vacuum states.
Instantons can be thought of as describing quantum
tunneling transition processes (in Euclidean time) between two vacuum states characterized by different winding numbers $n$ and $n^\prime$. The topological charge carried by the
instanton exactly expresses the variation of the vacuum winding number for the
process: $Q=n-n^\prime$.

In real Minkwoski time, there is another interesting
class of solutions associated to topology-changing vacuum transitions known as
\emph{sphalerons}. Unlike instantons, sphalerons are saddle-point solutions of
the classical equations of motion of the Minkwoski-time Yang--Mills
theory~\cite{Christ:1979zm,Klinkhamer:1984di,Klinkhamer:2017fqi}. Sphalerons
are interpreted as describing transitions \emph{above} the energy barriers
separating the vacua with different winding number due to thermal fluctuations,
as opposed to instantonic quantum tunneling \emph{through} the barriers.
Sphalerons thus describe a very different kind of transitions compared to
instantons. As a matter of fact, according to this
interpretation, instantons are expected to be suppressed at very high
temperatures~\cite{tHooft:1976snw}, while for sphalerons the expectation is
that they will be thermally enhanced~\cite{Christ:1979zm,Klinkhamer:1984di}.

An important physical observable related to sphaleron transitions in Yang--Mills theories is
the rate of sphaleron transition processes per unit time and spatial volume,
also known as the \emph{sphaleron rate} $\Gamma_{\S}$:
\beq\label{eq:sphal_rate_def_real_time}
\Gamma_{\S} \equiv \lim_{t\to\infty}\lim_{L\to\infty} \frac{1}{t L^3} \Bigg\langle \left[\int_{-t/2}^{+t/2} \dd t^\prime \int \dd^3 x \, q(\vec{x},t^\prime)\right]^2 \Bigg\rangle_T = \int \dd t \, \dd^3 x \, \braket{q(\vec{x},t)q(\vec{0},0)}_T.
\eeq
The notation $\braket{\dots}_T$ in Eq.~\eqref{eq:sphal_rate_def_real_time} stands for the thermal expectation value:
\beq
\braket{q(\vec{x},t)q(\vec{0},0)}_{T} \equiv \frac{1}{\Tr\left\{\ee^{-\mathcal{H}_{\subQCD}/T}\right\}}\Tr\left\{\ee^{-\mathcal{H}_{\subQCD}/T} \, q(\vec{x},t)q(\vec{0},0)\right\}.
\eeq
Apart from its intrinsic theoretical significance, the sphaleron rate is an
interesting observable that plays a key role in several physical contexts. In
the context of heavy-ion collisions, a sphaleron transition occurring in a
magnetized strongly-interacting quark-gluon medium in the plasma
phase~\cite{Busza:2018rrf} is expected to lead to the phenomenon known as
\emph{Chiral Magnetic Effect}
(CME)~\cite{Kharzeev:2007jp,Fukushima:2008xe,Fukushima:2010vw,Kharzeev:2013ffa}.The
creation of topological excitations via sphaleron transitions  will lead, by
virtue of the chiral anomaly, to a \emph{chiral imbalance}, i.e., an excess of
left-handed or right-handed fermion zero-modes. Since the spin of both
left-handed and right-handed modes will align in the same direction of the
magnetic field, their momenta will be instead aligned in opposite directions
with respect to the magnetic field, eventually leading to a net electric
current flowing in the plasma along the magnetic field. This process is
expected to take place out-of-equilibrium as, at $\theta=0$, chiral imbalances
must average to zero in thermal equilibrium. The sphaleron rate exactly
describes the equilibration process of the axial quark number density
$\rho_{5}$ in the presence of $n_{\subf}$ light
flavors~\cite{Moore:2010jd,BarrosoMancha:2022mbj}:
\beq
\frac{\dd \rho_{5}}{\dd t} = -\frac{n_{\subf}}{T^3} \Gamma_{\S} \, \rho_{5} \, .
\eeq
The experimental detection of the CME is one of the major goals pursued at heavy-ion colliders~\cite{Feng:2025yte}, and precise theoretical inputs about $\Gamma_\S$ have been recognized to be of the utmost importance for this quest since a decade~\cite{Kharzeev:2024zzm}.

In more recent times, the sphaleron rate and its generalization to non-zero energy and momentum, the so-called topological rate:
\beq\label{eq:topological_rate_nonzeromomenta}
\Gamma_{\top}(E,p) = \int \dd^3 x \, \dd t \, \ee^{\ii p^\mu x_\mu} \, \braket{q(\vec{x},t)q(\vec{0},0)}_T, \quad \Gamma_\S = \Gamma_{\top}(0,0)\, , \quad p^\mu = \left(E, \vec{p}\right), \quad p = \vert \vec{p} \vert\, ,
\eeq
have been also identified as key quantities for the thermal production mechanism of axions in the hot early Universe~\cite{Notari:2022ffe,Bianchini:2023ubu,OHare:2024nmr,Bouzoud:2026rur}
(which is independent of the
misalignment mechanism previously mentioned in relation with the topological
susceptibility). The idea is that at very high temperatures real-time
topological transitions are effective in creating axions via the axion-gluon coupling $\propto a(x)q(x)$. On general grounds, this process could have happened out of thermal equilibrium, thus the momentum-dependent axion distribution function $f_p$ should be computed from the resolution of a Boltzmann equation~\cite{Notari:2022ffe}:
\beq
\frac{\dd f_p}{\dd t} = \left(1+f_p\right) \Gamma^{\pp}_{p} - f_{p} \Gamma^{\mm}_{p}.
\eeq
Here, the axion creation/annihilation rates $\Gamma^{\pp}_p/\Gamma^{\mm}_p$ can be related to the topological rate $\Gamma_\top$ as follows:
\beq
\Gamma^{\mm}_p = \ee^{E_p/T} \Gamma^{\pp}_p = \frac{1}{2E_p f_a} \Gamma_{\top}(E_p,p)\, , \qquad E_p = \sqrt{m_a^2 + p^2}\, .
\eeq
Again, we see that an important physical quantity that characterizes axion dynamics requires non-trivial inputs from QCD. This, in general, requires to resort to lattice QCD numerical methods to provide a reliable answer from first principles and in a fully non-perturbative fashion.

\section{Analytical and model predictions}\label{sec:analytic_res}

This section is devoted to a compact summary of the main analytic and model
predictions about the $\theta$-dependence of the free energy in QCD.
For asymptotically high temperatures the Dilute Instanton Gas Approximation (semiclassical methods supplemented by 
perturbation theory) is expected to be reliable.
In the low-temperature regime of QCD, characterized by confinement and spontaneous chiral symmetry
breaking, the large-$N$ $1/N$ expansion and Chiral Perturbation Theory (chiral effective Lagrangian approach) 
can instead be exploited.

\subsection{\texorpdfstring{$\theta$}{theta}-dependence from the Dilute Instanton Gas}
\label{subsect:DIGA}

The simplest scheme to obtain analytic results about $F(T,\theta)$ is the
Dilute Instanton Gas Approximation (DIGA). The main idea is to assume the QCD
functional integral to be dominated by a gas of
non-interacting pseudo-particles carrying a $Q=\pm 1$ topological charge
(instantons/anti-instantons, with all instantons assumed to be
indistinguishable, and analogously for the anti-instantons). In this
approximation, a generic configuration with charge $Q$ is obtained simply as a
superposition of $n_{\subI}$ instantons ($Q=1$) and $n_{\subA}$ anti-instantons
($Q=-1$) such that $Q=n_{\subI} - n_{\subA}$. The assumption of no interactions
is crucial to be able to write the partition function restricted to a given
topological simply in terms of a sum of products of single-particle partition
functions:
\beq
Z_n(T) = \sum_{n_{\subI}-n_{\subA}=n} \frac{1}{n_{\subI}!}\frac{1}{n_{\subI}!} Z_1(T)^{n_{\subI}}Z_1(T)^{n_{\subA}}.
\eeq
Here the factorials in the denominator are needed due to the assumption of
identical particles, while $Z_1(T) = Z_{\subI}(T) =Z_{\subA}(T)$ is the
one-particle partition function, i.e., the result of the integration over the
topologically trivial fluctuations and the instanton parameters (size, position, color
orientation). Physically, the diluteness assumption is equivalent to assume
the typical distance among such pseudo-particles to be much larger then their
typical size. In the end, one obtains for the partition function the following expression:
\beq
Z(T,\theta) = \sum_{n_{\subI}\,=\,0}^{\infty} \sum_{n_{\subA}\,=\,0}^{\infty}
\frac{1}{n_{\subI}!n_{\subA}!}Z_1^{n_{\subI}} Z_1^{n_{\subA}}\ee^{\ii\theta(n_{\subI}-n_{\subA})} = \left(\sum_{n_{\subI}\,=\,0}^{\infty} \frac{Z_1^{n_{\subI}}}{n_{\subI}!}\ee^{\ii \theta n_{\subI}}\right) \left(\sum_{n_{\subA}\,=\,0}^{\infty} \frac{Z_1^{n_{\subA}}}{n_{\subA}!}\ee^{-\ii \theta n_{\subA}}\right)
= \exp\left\{\ee^{\ii\theta}Z_1\right\} \exp\left\{\ee^{-\ii\theta}Z_1\right\}
= \exp\left\{2\cos(\theta)\, Z_1\right\}.
\eeq
The $\theta$-dependent part of the free energy thus reads ($V=T/L^3$, as before):
\beq\label{eq:free_energy_theta_DIGA}
F(T,\theta)= \frac{2Z_1(T)}{V} \left[1-\cos(\theta)\right] = \chi(T)\left[1-\cos(\theta)\right]
\, .
\eeq
The diluteness hypothesis is thus strong
enough to fix completely the $\theta$-dependence of the free energy up to an
overall pre-factor, $\chi(T)$, which remains unknown. In particular, matching
to Eq.~\eqref{eq:thetadep_free_energy_general}, the higher-order coefficients are given by:
\beq
\label{eq:b2n_DIGA}
b_{2n} = \frac{2(-1)^n}{(2n+2)!} \, , \qquad b_2 = -\frac{1}{12} \simeq -0.08333 \,, \qquad b_4 = \frac{1}{360} \simeq 2.78 \times 10^{-3} \,,\dots
\eeq
all turn out to be temperature-independent and finite.

The evaluation of $\chi(T)$ requires some extra assumptions, as one needs to
evaluate $Z_1(t)$. A possible way to do it is via a saddle-point approximation,
assuming small fluctuations around an instanton configuration, so that:
$\mathcal{S}_{\subE}^{\supYM} = \mathcal{S}_{\subI} + \frac{\delta^2
\mathcal{S}_{\subE}}{\delta A^2} \delta A^2$, with $\mathcal{S}_{\subI} =
\mathcal{S}_{\min} = \frac{8
\pi}{g^2}$~\cite{Gross:1980br,Pisarski:1980md,Boccaletti:2020mxu}. This leads
to a Gaussian functional integral whose calculation can be carried on. In this
approximation, the running coupling in $\mathcal{S}_{\subI}$ should be
evaluated at the instanton size scale $\mu = 1/\rho$. It is thus clear that the
only possibility for this semiclassical approximation to be reliable is to
require that the path integral is dominated by small-size instantons, so that
$g(\rho) \ll 1$. However, this is not the case in general, as the instanton
size distribution is found to be
\beq
d_{\subI}(\rho) \propto \rho^{\frac{11}{3}N - \frac{2}{3}N_{\subf} - 5}
\eeq
which diverges for large $\rho$ when $N \geq 3$ and $N_{\subf} < 9$. Therefore,
in this form, the DIGA is simply internally inconsistent: it assumes the path
integral to be dominated by small weakly-interacting instantons, but is actually dominated by
large strongly-interacting ones.

However, this scenario drastically changes when the system is studied for
asymptotically high temperatures $T \gg \Lambda_{\subQCD}$. In this case, a
sort of Debye-screening mechanism suppresses large instanton with size $\rho
\gg 1/T$. In this case, one expects the DIGA model to work, although \emph{a
priori} it is not clear how large $T$ must be for this to be true. 

By performing the integration on the field fluctuations and the instanton
parameters we find, for large values of $T$ and in the presence of $N_{\subf}$
light degenerate quarks with mass $m$~\cite{Gross:1980br,Pisarski:1980md,Boccaletti:2020mxu}:
\beq
\chi(T) \propto m^{N_{\subf}} T^{4 - \frac{11}{3} N  - \frac{1}{3}N_{\subf}} \, .
\label{eq:asym_chi_DIGA}
\eeq
The DIGA prediction is thus that $\chi(T)$ is suppressed in the chiral limit as
$\chi \sim m^{N_{\subf}}$, eventually leading to the vanishing of $\chi$ (and
thus to no $\theta$-dependence) for exactly massless quarks. Moreover,
$\chi(T)$ is also strongly suppressed with the temperature: as an example, for $N=N_{\subf}=3$ one finds $\chi \sim T^{-8}$. As a
conclusive comment, we stress that the diluteness hypothesis and the 1-loop
approximation employed for the estimation of $\chi$ are not bound to become
reliable in the same temperature range, as they are based on rather different
assumptions~\footnote{If the perturbative computation is reliable then
$\chi(T)\ll 1$ and instantons are diluted, however there is no reason for the inverse
implication to be true in general.}. Thus, they could very well set in at different temperatures. As we shall discuss later, this is exactly the observed behavior from numerical studies.

\subsection{\texorpdfstring{$\theta$}{theta}-dependence in the \texorpdfstring{large-$N$}{large-N} limit}

We have discussed in Sec.~\ref{sec:WittenVeneziano} how assuming a non-trivial
$\theta$-dependence of QCD in the large-$N$ limit allows to derive a
phenomenological estimate of the topological susceptibility. Using further
theoretical arguments, one can actually go beyond and obtain some qualitative
predictions for the whole $\theta$-dependence of the free energy. In
particular, in the large-$N$ limit \emph{\'a la} 't Hooft, one expects the
following general property to hold~\cite{Witten:1980sp,Witten:1998uka}:
\beq
F(T,\theta,N) \underset{N\to\infty}{\sim} N^2 \bar{F}(T,\bar{\theta}), \qquad \bar{\theta}= \frac{\theta}{N}.
\eeq
The $\sim N^2$ scaling of the free energy simply descends from its extensive
property, which leads one to argue that, at large $N$, $F$ should be
proportional to the number of gluonic degrees of freedom (which are $N^2-1$ for
$\SU(N)$ gauge group). Instead, the fact that the $\theta$ and $1/N$ dependence
should appear only in the combination
$\bar{\theta}$ stems from the following rescaling of the Yang--Mills action in
the presence of a $\theta$-term:
\beq
\mathcal{S}^{\supYM}_{\subE}(\theta) = \frac{N}{\lambda} \left[\frac{1}{4} \int \dd^4 x \, G^a_{\mu\nu}(x)G^a_{\mu\nu}(x) -\ii \lambda\frac{\theta}{N} Q\right] \, .
\eeq
Matching to Eq.~\eqref{eq:thetadep_free_energy_general}, and 
imposing $\chi\sim\mathcal{O}(N^0)$ we find:
\beq
\bar{F}(T, \bar \theta) 
\underset{N\,\to\,\infty}{\sim} \frac{1}{2}\chi(T) \bar{\theta}^2\left(1+\sum_{n\,=\,1}^{\infty} \bar{b}_{2n}(T) \bar{\theta}^{2n}\right) \, , \qquad \qquad  \bar{b}_{2n}(T) = b_{2n}(T) N^{2n} \sim \mathcal{O}(N^0) \, .
\eeq
Using the assumption $\chi\sim\mathcal{O}(N^0)$ (consistent with the
Witten--Veneziano formula) and large-$N$ scaling, we thus predict the $b_{2n}$
coefficient to vanish in the large-$N$ limit as $1/N^{2n}$~\footnote{The same conclusions can
also be reached from a large-$N$ analysis of QCD $\theta$-dependence
obtained from chiral effective
theories~\cite{Vonk:2019kwv,GomezNicola:2019myi}.}.
Note that this result is in sharp contradiction with DIGA expectations~\eqref{eq:b2n_DIGA}, 
however DIGA predictions are reliable only in the high-temperature limit, while the large-$N$ expansion 
is expected to be reliable (due to the large-$N$ factorization) only in the low temperature regime.

Unfortunately, in Yang--Mills theories it not possible to obtain more
quantitative information regarding the $b_{2n}$ coefficients, as it happens instead in
lower-dimensional QCD-like vector models exhibiting a non-trivial
$\theta$-dependence, such as the $2d$ $\CP^{N-1}$ models, for which the $1/N$ expansion can be carried out analytically~\cite{DAdda:1978vbw, Luscher:1978rn,
Witten:1978bc, Campostrini:1991kv, Campostrini:1991tw, DelDebbio:2006yuf,
Rossi:2016uce, Bonati:2016tvi, Sugeno:2025exv}. There is however an important
conclusion that can be drawn already based on the results here outlined. At
first glance, one may wonder how is it possible that a function that is
expected to be $2\pi$ periodic in $\theta$ for all $N$ values can actually be a
function of $\bar{\theta}=\theta/N$ in the large-$N$ limit.  The only
possibility is that of a non-analytic behavior, where level-crossings of
different branches take place, leading in the end to a function $\bar{F}(T,\bar{\theta})$
with the correct periodicity. Recalling that only the leading
$\mathcal{O}(\theta^2)$ term survives at large $N$, one expects~\cite{Witten:1980sp,Witten:1998uka}:
\beq
\bar{F}(T,\bar{\theta}) = \frac{1}{2}\chi(T) \, \min_{k \, \in \, \mathbb{Z}} \left(\bar{\theta} + \frac{2\pi k}{N}\right)^2 = \frac{1}{2}\chi(T) \, \min_{k \, \in \, \mathbb{Z}} \left(\frac{\theta + 2\pi k}{N}\right)^2 \, .
\eeq
This function exhibits cusps in $\theta=\pi + 2\pi k$. At these points of non-analyticity, provided that $\chi(T) \ne 0$, the CP symmetry is spontaneously broken due to the co-existence of degenerate vacua with equal free energies. Note that this fact is, again, at odds with DIGA predictions at high temperature, which instead predicts a smooth behavior in $\theta=\pi$.

\subsection{\texorpdfstring{$\theta$}{theta}-dependence from Chiral Perturbation Theory}\label{sec:chiPT}

In the presence of light quark flavors, one can describe the low-temperature regime of QCD via chiral effective Lagrangians. The construction of the effective low-energy
theory for QCD is achieved by writing down an effective Lagrangian implementing
the expected chiral-symmetry breaking pattern in terms of the lightest degrees
of freedom of the theory, i.e., the pions assuming the up and down quarks to be
light, or the lightest octet (pions,kaons and $\eta$), assuming the up, down
and strange quarks to be light.
This framework, known as Chiral Perturbation Theory ($\chi$PT), can be
systematized as an expansion in powers of $\mathcal{O}(p^2)$ (i.e., in the
number of the field derivatives) and $\mathcal{O}(m)$, with $m$ generically
indicating a light quark mass.

At the lowest order in $p^2$ and $m$, and for $N_{\subf}=2$ light flavors, the $\chi$PT Lagrangian reads:
\beq
\mathcal{L}_{\eff}(\theta) = \frac{F_\pi^2}{4}\Tr\left[\partial_\mu U^{\dag} \partial^\mu U\right] + \frac{\Sigma}{2}\Tr\left[\mathcal{M}(\theta) U + \mathcal{M}^\dag(\theta)U^\dag\right] \, , \qquad \mathcal{M}(\theta) \equiv \ee^{-\ii \theta/2} \mathrm{diag}(m_u,m_d) \, ,
\eeq
where $U=\exp(\ii\pi_a\sigma_a/F_\pi)$, and the field $\pi_a$ is the
Goldstone-boson field describing pions, $F_\pi = \lim_{m_u,m_d\to 0}
\frac{1}{\sqrt{2}m_\pi}\bra{0} \overline{\psi}\gamma_0\gamma_5\psi
\ket{\pi,\vec{p}=\vec{0}}\simeq 87\text{
MeV}$~\cite{FlavourLatticeAveragingGroupFLAG:2021npn} is the pion decay
constant in the chiral limit, and $\Sigma = - \lim_{m_u,m_d\to
0}\braket{\overline{\psi}\psi} \simeq (272\text{ MeV})^3 \simeq 2 \times
10^{-2} \mathrm{GeV}^3$~\cite{FlavourLatticeAveragingGroupFLAG:2021npn} is the
quark chiral condensate in the chiral limit. These two low-energy constants, together with the quark masses, are enough to describe the $\chi$PT theory at leading order. Finally, the quark mass term $\mathcal{M}(\theta)$ features a complex phase including the $\theta$ term,
akin to the quark mass matrix in the full theory after a U(1) axial rotation.

At exactly $T = 0$, the free energy receives contribution only from the ground
state energy, $F(T=0,\theta)\equiv E_{\subG}(\theta)$. This is given by the
minimum value of the potential for the field $U$ of
$\mathcal{L}_{\eff}(\theta)$, namely, $\mathcal{V}_{\eff}(\theta)=\Sigma \,
\Re\left[\mathcal{M}(\theta)U_{\min}\right]$. The minimization of
$\mathcal{V}_{\eff}(\theta)$ leads to~\cite{DiVecchia:1980yfw} (see also
Refs.~\cite{GrillidiCortona:2015jxo,Luciano:2018pbj}):
\beq
E_{\subG}(\theta) = - F_\pi^2 \left[ m_\pi^2(\theta) - m_\pi^2 \right] \, .
\label{eq:f_chiPT}
\eeq
Here, $m_\pi(\theta)$ and $m_\pi$ denote, respectively, the $\theta$-dependent and the $\theta=0$ pion masses:
\beq
m_\pi^2(\theta) = m_\pi^2\sqrt{1-\frac{4z}{1+z}\sin^2\left(\frac{\theta}{2}\right)}, \qquad m_\pi^2 = \frac{\Sigma}{F_\pi^2}(1+z)m_u, \qquad z=\frac{m_u}{m_d} .
\label{eq:mpi_chiPT}
\eeq
Matching Eq.~\eqref{eq:f_chiPT} to the general parameterization in
Eq.~\eqref{eq:thetadep_free_energy_general} one finds, for example:
\beq
\chi = \frac{z}{1+z}m_\pi^2 F_\pi^2 = \Sigma \left(m_u^{-1} + m_d^{-1}\right)^{-1} \, , \qquad b_2 = -\frac{1}{12}\frac{1+z^3}{(1+z)^3} \, .
\label{eq:nf2_leading_chiPT}
\eeq
Interestingly, we see that the chiral suppression of $\chi$ is milder than in
the DIGA case: $\chi \sim m$, as opposed to $\chi(T)\sim m^2$ for an equal
number of light quark flavors. The $b_{2n}$ coefficients instead only depend on
the up/down quark mass ratio $z$ and are thus finite in the chiral limit.

This computation can be extended in several ways. At leading order, one can
easily take into account also a third quark flavor, obtaining for the susceptibility the following expression: $\chi = \Sigma (m_u^{-1} +m_d^{-1} +m_s^{-1})^{-1}$. It
is instead much more involved to derive an expression at next-to-leading-order
in $\chi$PT due to the appearance of further additional low-energy constants. A
full analytic expression is available for the 2-flavor
case~\cite{Guo:2015oxa,GrillidiCortona:2015jxo,Bonati:2015vqz,GomezNicola:2019myi,Lu:2026fhg},
while no closed form exists in the 3-flavor
one~\cite{Guo:2015oxa,Lu:2020rhp}. Focusing just on the first three terms in
the $\theta$ expansion of the $N_{\subf}=2$ case, and considering only the two
interesting cases $z=1$ (isospin symmetric case) and $z=0.48$ (physical up/down
ratio with isospin breaking) one obtains:
\begin{align}
\chi &= \left[75.5(5)~\mathrm{MeV}\right]^4, & b_2 &= -0.29(2), & b_4 &= -0.00028(7), & z &=0.48 \, ,\\
\chi &= \left[77.8(4)~\mathrm{MeV}\right]^4, & b_2 &= -0.22(1), & b_4 &= -0.00017(7), & z &=1 \, .
\end{align}
In the case of $\chi$, it was shown that next-to-next-to-leading-order $\chi$PT corrections are small and accidentally canceled by the
$O(\alpha_{\rm em})$ effects due to QED, thus these estimates remain essentially unaltered~\cite{Gorghetto:2018ocs}.  Finally,
one can extend the analysis also to the finite temperature case. Due to the assumptions behind $\chi$PT, it is natural to believe that such predictions should break down close to the QCD chiral crossover temperature above which chiral symmetry is approximately restored, $T_c\simeq 155$ MeV. In general, $\chi$PT predicts a slow temperature-suppression of $\theta$-dependence. For the first two coefficients, one approximately has $\chi(T)/\chi(0) \simeq 1 -\mathcal{O}(T^2)$ and $b_2(T)/b_2(0) \simeq 1 - \mathcal{O}(T^2)$, with these leading-order finite-$T$ corrections suppressed as $\sim\mathcal{O}(1/N^2)$. More details can be found in Refs.~\cite{GrillidiCortona:2015jxo,Lu:2026fhg}.

\section{Lattice results}\label{sec:num_res}

So far, we have seen how analytical model calculations allow to characterize $\theta$-dependence in QCD in certain regimes (low and high temperatures). In general, one has to resort to numerical Monte Carlo simulations of lattice QCD in order to be able to study $\theta$-dependence from first principles and in a fully non-perturbative fashion. This is however in practice a highly non-trivial task, as several computational challenges have to be faced.

\subsection{The lattice approach in a nuthsell}
\label{subsec:lattice}

Let us start by briefly reviewing the fundamentals of the lattice approach. The lattice formulation relies on the fact that, in Euclidean time, the factor $\ee^{-\mathcal{S}_{\subE}}$ appearing in the functional integral can be treated as a Boltzmann-like probability distribution for the fields. Once the continuum finite-volume space time is substituted with a finite space-time lattice with lattice spacing $a$, the path integral reduces to an ordinary integral that can be efficiently computed stochastically via Monte Carlo importance sampling methods. The lattice spacing $a$ also introduces a UV cut-off $\Lambda_{\scriptscriptstyle{\rm UV}} \sim 1/a$ that acts as a regulator for the Quantum Field Theory. The cut-off is removed in the continuum limit $a\to 0$, when the spacing of the lattice grid shrinks to zero, and the lattice space-time tends to the physical continuum space-time. The continuum limit is practically approached, by virtue of asymptotic freedom, by making the bare coupling of the lattice theory smaller and smaller.

In the lattice formulation adopted in numerical simulations, only the
gauge fields appear explicitly, as quark fields are exactly integrated out (being the fermion action quadratic in the spinor fields). More precisely, a generic expectation value takes the form:
\beq
\braket{\mathcal{O}} = \int [\dd U] \frac{1}{Z_{\subL}}\exp\left\{-S_{\subYM}^{\supL}[U]\right\} \prod_{\mathrm{f} \, = \, 1}^{N_{\subf}}\det\left\{D[U]+m_{\subf}\right\} \mathcal{O}[U] \equiv \int [\dd U] \, P[U] O[U] \, ,
\eeq
where the dynamical variables are the so called gauge links $U_\mu$, which are $\SU(N)$ matrices that represent the elementary parallel transport between the lattice sites $x$ and $x+a\hat{\mu}$. Sufficiently close to the continuum limit $a\to 0$, one approximately has $U_\mu(x)\simeq \ee^{\ii a A_\mu(x)}$, with $A_\mu(x)$ the continuum gluon field. The Dirac determinant, instead, encodes quark dynamics and is the result of the integration over fermion fields. Finally, $[\dd U]$ represents the gauge-invariant $\SU(N)$ Haar measure for each link of the lattice.

Once a sample of the gauge links $\{U_i\}$ has been generated from the
Monte Carlo by sampling $P[U]$, one evaluates $\mathcal{O}_i=\mathcal{O}[U_i]$, where $O$ is an observable, and estimates $\braket{\mathcal{O}}$ from the sample average of $\{\mathcal{O}_i\}$. This estimate of $\braket{O}$ will be affected by a statistical error which will decrease with the square root of the sample size. Gluonic observables are typically expressed as gauge-invariant traces of products of links around closed paths. The simplest example is the \emph{plaquette} $U_{\mu\nu}(x)=U_{\mu}(x) U_{\nu}(x+a\hat{\mu}) U^\dagger_{\mu}(x+a\hat{\nu}) U^\dagger_{\nu}(x)$, which is the simplest planar path one can define on an hypercubic lattice, and it represents the most straightforward discretization of the gauge field strength $G_{\mu\nu}(x)$. On the other hand, fermionic observables can always be expressed in terms of traces involving the inverse of the discretized Dirac
operator $D[U]$ (with $D^{-1}$ representing the quark propagator on the
lattice). Several possible different lattice Dirac operators can be defined.
The most commonly employed formulations (the so-called Wilson and staggered
formulations) explicitly break the chiral symmetry at finite lattice spacing.
Discretizations with better chiral properties~\cite{Ginsparg:1981bj}  are the so-called domain
wall fermions~\cite{Kaplan:1992bt,Shamir:1993zy,Furman:1994ky} and overlap fermions~\cite{Neuberger:1997fp,Neuberger:1998wv,Luscher:1998pqa}, which are however much more demanding from the computational point of view than Wilson and staggered fermions.

\subsection{Theoretical and computational challenges in lattice studies of Yang--Mills topology}

Studying topological quantities require to address a series of non-trivial
issues. We briefly recall here a few very general ones (see Ref.~\cite{Vicari:2008jw} for more details), leaving for later the description of other more specific problems.

A first theoretical difficulty already emerges when trying to define a
topological charge for the lattice theory. Indeed, strictly speaking topology
is a feature of the continuum fields and is not well defined on a space-time
lattice. This is in principle not an obstacle, as, eventually, a proper notion of homotopy classes of gauge fields will be recovered in the continuum limit.
However, this aspect signals that the definition of the topological charge on the lattice requires some care. 

The simplest approach consists in writing a discretized topological
charge density in terms of the product of two plaquettes (or larger loops) to
reproduce $G\widetilde{G}$ (gluonic definitions) in the continuum limit. Such
\emph{field-theoretic} definitions will not give rise to an integer topological
charge due to multiplicative
renormalizations~\cite{Campostrini:1988cy,Campostrini:1989dh}, and further
divergent additive renormalizations will show up when using them to define the
cumulants of the topological charge distribution due to the appearance of
contact terms~\cite{DiVecchia:1981aev,DiVecchia:1981hh,
Campostrini:1989dh,DElia:2003zne}.
Alternatively, \emph{geometric} definitions of $Q$
can be introduced, which are exactly integer even at finite lattice spacing.
These class of discretized charges are defined by interpolating lattice fields
to obtain pseudo-continuum ones, where a winding number can be
defined~\cite{Berg:1981er, Luscher:1981zq, Woit:1983tq, Phillips:1986qd}. These
definitions however are plagued by the presence of the so-called
\emph{dislocations}: lattice configurations with a large action due to UV
fluctuations at the scale of the lattice spacing, for which the interpolation is
highly ambiguous. For entropic reasons, dislocations can proliferate as the
continuum limit is approached, obscuring the correct behavior of the
cumulants of the topological charge distribution~\cite{Berg:1981nw,
Luscher:1981tq}. 

In both cases, it is clear that the origin of the problem is a UV contamination
of the lattice topological charge operator due to the finite cut-off set by the
inverse lattice spacing. An effective way to get rid of such contamination is
to smoothen gauge fields by some smoothing procedure. Smoothing methods are
typically iterative algorithms that remove UV noise by driving the
configuration towards the minimum of the lattice action. Removed fluctuations
stay below a new scale introduced by smoothing, the smoothing radius $R_s$,
which is proportional to the square root of the number of performed steps of
the smoothing algorithm. Several smoothing techniques have been proposed and
employed, such as cooling~\cite{Berg:1981nw, Iwasaki:1983bv, Teper:1985rb,
Ilgenfritz:1985dz, Campostrini:1989dh} and gradient
flow~\cite{Narayanan:2006rf,Luscher:2009eq, Luscher:2010iy, Luscher:2011bx}.
Different smoothing methods are naturally related to each
other~\cite{Alles:2000sc,Bruckmann:2006wf}, indeed all these
techniques have been shown to give compatible results when the
smoothing radii are properly matched to each
other~\cite{Bonati:2014tqa,Alexandrou:2017hqw}.

There is another broad class of possible discretizations of the topological
charge rooted on the index theorem (fermionic definitions). On the lattice,
when using non-chiral formulations of the Dirac operator, a definition
involving only the sum over the chirality of the zero-modes is impossible, as
no exact zero-mode is present in the spectrum of the discretized Dirac
operator. In this case, the strategy is to extend the sum to non-zero modes up
to a certain cut-off scale $M$~\cite{Luscher:2004fu,Giusti:2008vb,
Luscher:2010ik, Bonanno:2019xhg}, which will act similarly to the smoothing
radius $R_s$ in gluonic definitions, i.e., by cutting away the UV modes.
Remarkably, a similar approach, based on the filtering of Dirac eigenmodes, was suggested originally as an alternative smoothing technique for the gauge field themselves~\cite{Gattringer:2002gn}. In this approach,
multiplicative renormalizations are still present~\cite{Smit:1987zh,Smit:1987jh} and have to be explicitly taken into
account, but no additive divergent renormalization is present. If instead
chiral lattice fermions are employed, one recovers an integer valued definition of the winding number~\cite{Giusti:2001xh,Giusti:2004qd}, simply obtained from the counting of left-handed and right-handed zero-modes. However, ambiguities may emerge also in this case, see, e.g.,
Refs.~\cite{DelDebbio:2003rn,DelDebbio:2004ns}. 

Apart from the theoretical problem of
defining the topological charge on the lattice, there is also a very serious
computational problem affecting standard Monte Carlo techniques when it comes
to their efficiency in sampling the topological charge distribution.  In
general, all Monte Carlo sampling algorithms are
affected by critical slowing down when approaching a critical point where the correlation length in lattice units diverges, 
as it happens when the continuum limit of a lattice gauge theory
emerges. This means, in practice, that the sampling becomes less and less efficient as
$a\to 0$, requiring more and more time to generate statistically independent samples. For lattice theories with non-trivial topological features, critical slowing down is particularly severe, and leads eventually to a phenomenon known as \emph{topological
freezing}~\cite{Alles:1996vn,DelDebbio:2002xa,DelDebbio:2004xh,
Schaefer:2010hu, Bonati:2017woi}, which, in a certain sense, is the bad side of
the fact that the very concept of topology is recovered in the continuum limit.
Indeed, when $a\to 0$, due to the emergence of infinitely-high potential energy
barriers separating the different homotopy classes, standard Monte Carlo
sampling methods become less and less efficient in making the topological
charge jump from a topological sector to another.  This eventually leads to a
freezing of the value of the topological charge in a fixed sector during the
Monte Carlo simulation. Clearly, this prevents a proper sampling of the
topological charge distribution $\mathcal{P}(Q)$, and thus of the calculation
of topological quantities such as the cumulants of $\mathcal{P}(Q)$.
Topological freezing is particularly severe at large-$N$, which is a regime
that has been extensively explored to assess the validity of the
Witten--Veneziano mechanism from first principles. In particular, as $N$ is
increased, freezing shows up at coarser and coarser lattice spacings, likely because the action gap between different topological sectors grows with $N$: this makes the study of the continuum limit of topological quantities towards $N\to \infty$ very hard.

A further computational issue emerges in the presence of light fermions. In the
continuum, one expects the fermion determinant to suppress gauge configurations
with non-zero topological charge, due to appearance of zero-modes in the Dirac
spectrum. If a non-chiral lattice discretization is employed, no exact
zero-mode appears, and would-be-zero modes can be quite large compared to the
light quark mass, thus spoiling the expected suppression in the continuum. As a
result, cut-off effects in the presence of light dynamical fermions are much
larger compared to those found in the pure-glue theory: this makes it
necessary, in order to obtain reliable continuum extrapolations, to use smaller
lattice spacings than usual, thus making the impact of topological freezing
even more disruptive.

\subsection{Lattice results for \texorpdfstring{$\theta$}{theta}-dependence at low temperatures: chiral and large-\texorpdfstring{$N$}{N} scaling}

Let us start our presentation of lattice results about $\theta$-dependence from
zero-temperature pure-glue simulations without dynamical fermions (meaning in
practice that the determinant is omitted from the functional integral). The
principal goals of these studies are to test the validity of the
Witten--Veneziano (WV) formula and, more generally, to validate from first
principles the predicted large-$N$ scaling for the $\theta$-dependence. Thus,
although the pure-glue theory is an ideal limit of real-world QCD, its
investigation is actually extremely relevant for strong interactions, both
theoretically and for phenomenology. 

\begin{figure}[!t]
\centering
\includegraphics[scale=0.485]{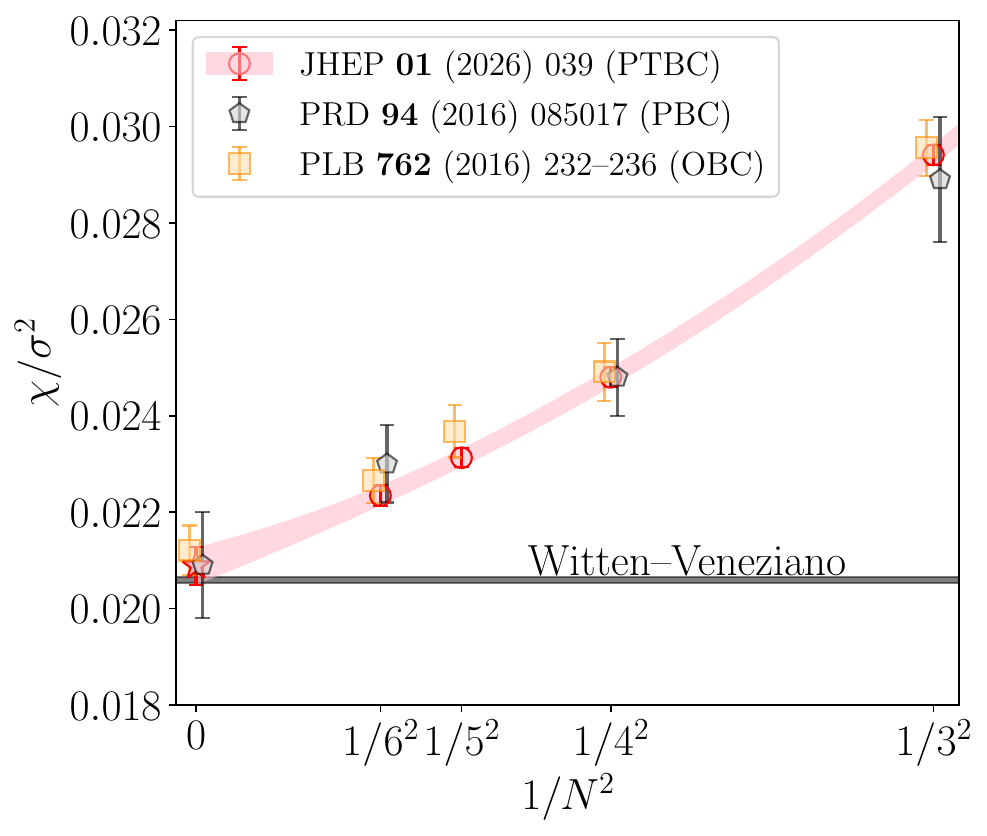}
\includegraphics[scale=0.48]{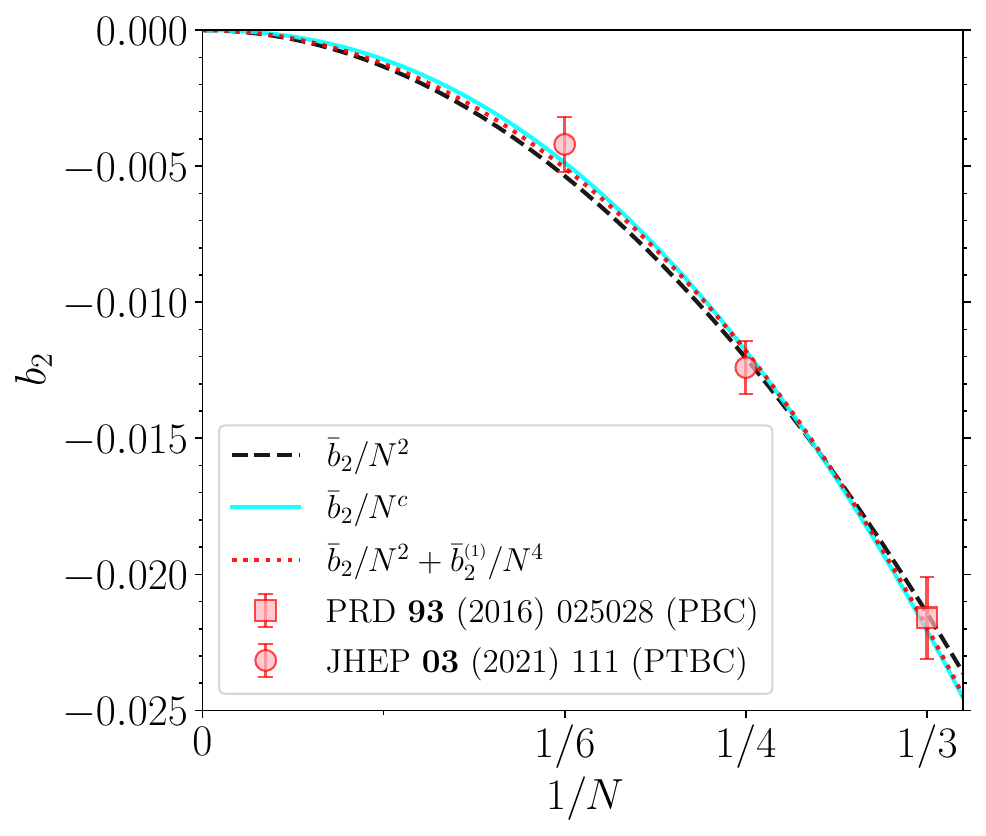}
\caption{Left: extrapolation towards $N\to\infty$ of $\chi/\sigma^2$, with
$\sigma$ the string tension~\cite{Bonati:2016tvi,Ce:2016awn,Bonanno:2025eeb}. Best fit yields $\chi/\sigma^2= 0.02088(39) +
0.044(12)/N^2+ 0.293(83)/N^4$ (source: Ref.~\cite{Bonanno:2025eeb}). Right:
Extrapolation toward $N\to\infty$ of $b_2$~\cite{Bonati:2015sqt,Bonanno:2020hht} (source:
Ref.~\cite{Bonanno:2020hht}). Best fit according to $b_2=\bar{b}_2/N^2$ yields
$b_2=-0.193(10)/N^2$ fixing the exponent $c = 2$ (dashed line). This result is stable
within errors if the exponent is left as a free parameter, $c=2.17(26)$ (solid
line), and when a further $1/N^4$ correction is added (dotted line).}
\label{fig:largeN}
\end{figure}

First attempts~\cite{DiVecchia:1981aev,DiVecchia:1981hh} to determine $\chi$
from the lattice were carried out soon after the first
pioneering lattice simulations~\cite{Creutz:1979dw, Creutz:1980zw},
however in these seminal works the multiplicative
renormalization~\cite{Campostrini:1988cy,Campostrini:1989dh} of the
field-theoretic discretization was not taken into account, thus obtaining a
value of $\chi$ much smaller than the one expected by using the
WV formula. After this aspect was taken
into account, results falling in the correct ballpark were found for the SU(2)
and SU(3) theory~\cite{Alles:1996nm,Alles:1997qe}. Later on, such
determinations have been constantly refined and cross-checked with very
different methods, both based on gluonic~\cite{Durr:2006ky, Ce:2015qha,
Bonati:2015sqt, Bonanno:2020hht,Athenodorou:2020ani, Athenodorou:2021qvs,
Bonanno:2023ple,Durr:2025qtq,Bonanno:2025eeb} and
fermionic~\cite{DelDebbio:2004ns,Luscher:2010ik,Cichy:2015jra, Bonanno:2019xhg}
discretizations of the topological charge, achieving eventually a
per-cent level control on the large-$N$ limit of $\chi$. This is illustrated in
Fig.~\ref{fig:largeN} (left panel), where the currently most accurate
determinations of $\chi$ in the large-$N$ limit are shown. These are obtained either using standard periodic boundary conditions (PBC), open boundary conditions in the time direction (OBC, see \cite{Luscher:2011kk}), or a combination of the two in a parallel tempering framework (Parallel Tempering on Boundary Conditions or PTBC, see \cite{Hasenbusch:2017unr}). The OBC and PTBC strategies have been instrumental to improve accuracy, as they allow to mitigate the severe topological freezing
affecting large-$N$ simulations. 

In more recent times the validity of
the WV picture was further supported by the first lattice computation of the
topological susceptibility slope $\chi^\prime$ of the pure SU(3) gauge
theory~\cite{Bonanno:2023ple}.
This computation is significantly more complex than that
of the topological susceptibility, because this observable does not depend just on the total charge, which is stable under
smoothing, but on the details of the local topological charge density.
Thus, in this case, smoothing impacts the actual value of the observable, and
the physical determination is obtained in the limit in which the smoothing
radius tends to zero. Eventually, it is found that the bound $\vert \chi^\prime
\vert \ll \chi/m_{\eta^\prime}^2$ is actually respected, as $\chi^\prime
m_{\eta^\prime}^2 / \chi \sim \mathcal{O}(10^{-1})$~\cite{Bonanno:2023ple}.

Another important aspect, in relation with the overall validity of the WV
picture, is the large-$N$ scaling of higher-order cumulants, parameterized
through the coefficients $b_{2n}$. On general grounds, their determination is
much harder than that of $\chi$: the ratios defined in Eq.~\eqref{eq:b2n_def} 
become noisier and noisier as the volume is increased because they parameterize
deviations of $\mathcal{P}(Q)$ from a Gaussian, which become harder and harder
to detect as $V\to\infty$, due to the central limit theorem. In practice, a
significant increase in statistics, proportional to (powers of) the volume, is
necessary to keep the statistical error of $b_{2n}$ constant. The
expected suppression of $b_{2n}$ for large $N$ makes this issue even worse. For
this reason, early determinations of $b_2$ for $N=3$ were affected by much
larger statistical errors compared to
$\chi$~\cite{DelDebbio:2002xa,DElia:2003zne,Giusti:2007tu}. A substantial
improvement, leading eventually to achieve a few-percent precision on $b_2$ as
a function of $N$, was obtained thanks to a novel methodology based on simulations performed at imaginary~\footnote{Note that $\theta$ has to be imaginary for the
path-integral weight $\exp\{-\mathcal{S}\}$ to be real and admit a
probabilistic interpretation, see Eq.~\eqref{eq:thetaterm}.} values of
$\theta$~\cite{Panagopoulos:2011rb}. Indeed, adding
a source term to the action allows to transfer the information on higher-order
cumulants at $\theta=0$ into lower-order cumulants at non-zero $\theta$ (this mechanism is very general, and it was was noted already in \cite{MBH86} when using an external magnetic field). The results obtained with this method~\cite{Panagopoulos:2011rb,Bonati:2015sqt,Bonati:2016tvi,Bonanno:2020hht}
allowed to prove the expected large-$N$ scaling $b_2 \sim 1/N^2$ with a
precision comparable to that of $\chi$, and the current most accurate
determinations are shown in the right panel of Fig.~\ref{fig:largeN}. Again,
let us also stress that the PTBC algorithm was instrumental to avoid topological
freezing and refine the statistical precision of $b_2$ data for $N>3$. Finally,
the determination of further $b_{2n}$ coefficients remains challenging still
nowadays. Currently, $b_4$ has been determined (with a $\sim 30\%$ accuracy)
only for the SU(2) theory~\cite{Bonanno:2018xtd}, and in this case it turns out
to be an order of magnitude smaller than $b_2$. This result is consistent with
existing upper bounds of $b_4$ for $N\ge3$~\cite{Bonati:2015sqt,Bonanno:2020hht} if we assume the scaling $b_{2n}\sim1/N^{2n}$ to be valid already starting from $N=2$.

\begin{figure}[!t]
\centering
\includegraphics[scale=0.485]{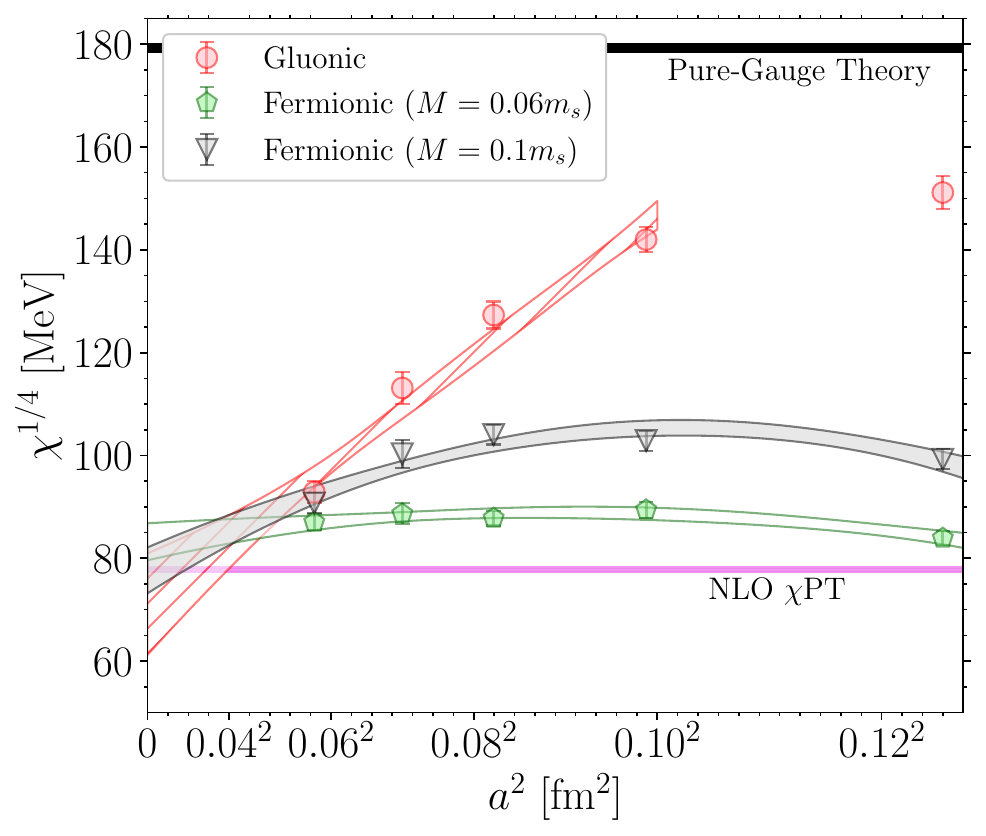}
\includegraphics[scale=0.48]{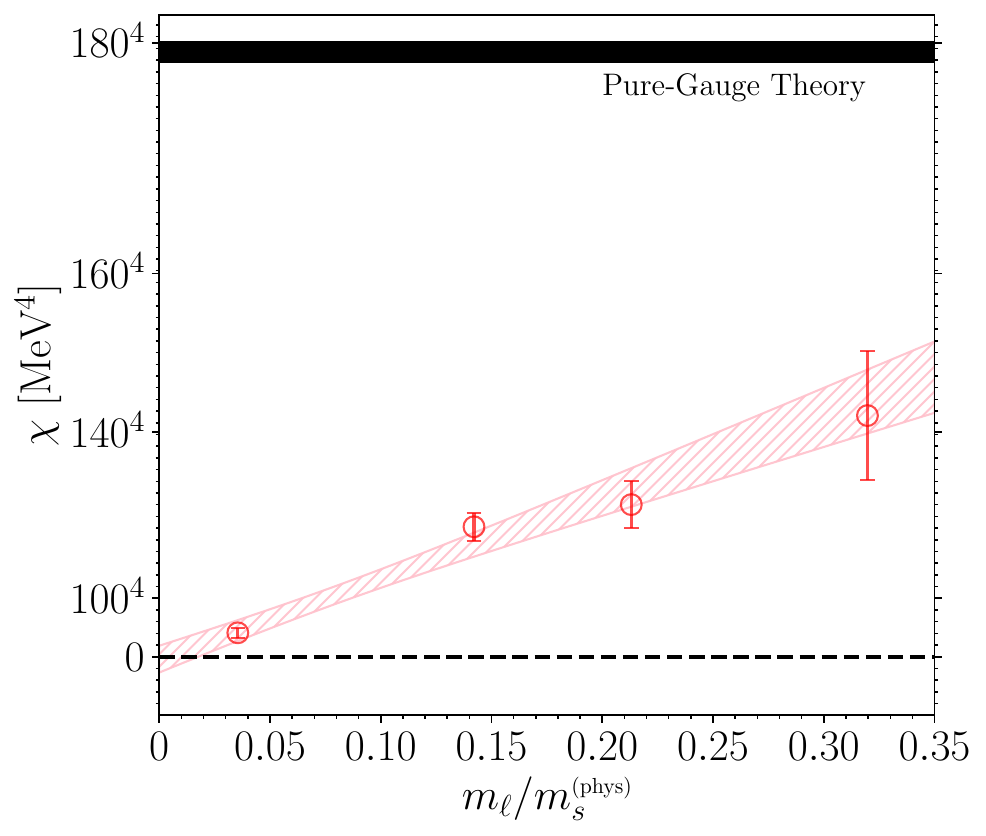}
\caption{Left: extrapolation towards the continuum limit of $\chi$ in $2+1$ QCD with physical quark masses with gluonic and fermionic discretization of the topological charge (source: Ref.~\cite{Athenodorou:2022aay}). Right: chiral extrapolation of the continuum extrapolations of $\chi$ in $2+1$ QCD obtained from a fermionic discretization (source: Ref.~\cite{Bonanno:2023xkg}). The x-axis reports the ratio of the degenerate light quark mass $m_{\ell}=m_u=m_d$ with respect to the physical strange quark mass (kept fixed as $m_\ell \to 0$). The pure-gauge value of $\chi^{1/4}$ comes from Ref.~\cite{Bonanno:2025eeb}.}
\label{fig:QCD}
\end{figure}

Let us now move to full QCD with dynamical quarks at the physical point,
starting from the the zero-temperature limit. Such an
investigation is primarily needed to verify that lattice simulations are able to reproduce the $\theta$-dependence predicted by $\chi$PT.
In the left panel of Fig.~\ref{fig:QCD} we report the continuum scaling of the
topological susceptibility with 2+1 flavor of dynamical fermions with physical
quark masses. In the lattice jargon, this means that the bare parameters of the
simulations are scaled so as to ensure that, while the lattice spacing is
changed, the common degenerate up/down quark mass $m_{\ell}\equiv m_u=m_d$ and
the strange quark mass are kept at their physical values. As it can be clearly
seen in Fig.~\ref{fig:QCD}, the gluonic definition of the topological
charge suffers for significant cut-off effects, i.e., it exhibits a rather
steep dependence on the lattice spacing as the continuum limit $a \to 0$ is
approached. In the same range of lattice spacings, the pure-gauge
susceptibility exhibits instead pretty mild cut-off effects. As a consequence,
the continuum extrapolation is much more difficult.  This was already
anticipated in the general discussion at the beginning of this section, and is
related to the presence of large (compared to the light quark mass) would-be
zero modes, that spoil the chiral suppression of the chiral determinant.
Although in the end the obtained continuum limit turns out to be in very good
agreement with the $\chi$PT prediction, it is reassuring to see that a
fermionic definition, suffering for much milder lattice artifacts, yields
compatible continuum extrapolations within errors. In the left panel of
Fig.~\ref{fig:QCD} we have reported two examples of results obtained by
using fermionic discretizations of the topological charge $Q$,
differing from each other for the value of the
cut-off $M$ used in the sum over the Dirac eigenmodes, see
Sec.~\ref{subsec:lattice}.  By now, several different studies of $\chi(T=0)$ in
full QCD exist, all giving consistent results in the continuum
limit~\cite{Bonati:2015vqz,Borsanyi:2016ksw,Alexandrou:2017bzk,Athenodorou:2022aay,Bonanno:2023xkg}.  

Finally, in the right panel of Fig.~\ref{fig:QCD} we report the chiral scaling
of continuum-extrapolated lattice determinations of $\chi$ as a function of the
light quark mass. The results reported in this plot
have been obtained adopting a fermionic definition
of $Q$, and keeping the strange quark mass fixed at its physical value
$m_s=m_s^{\scriptscriptstyle{(\rm phys)}}$ as $m_\ell$ is varied. The good
control achieved in the $T=0$ case on the continuum limit of the susceptibility
is shown by the fact that numerical data for
$\chi(m_\ell)$ approach zero as predicted by $\chi$PT in Eq.~\eqref{eq:nf2_leading_chiPT}: $\chi = \Sigma \, m_{\ell}/2$, with $\Sigma$ in very good agreement with other independent estimates of the chiral condensate~\cite{Bonanno:2023xkg}.

\subsection{Lattice results for \texorpdfstring{$\theta$}{theta}-dependence at high temperatures: across the phase transition}
\label{sec:HT}

As earlier discussed, $\chi$PT and large-$N$ predictions are well verified in
the low-temperature phase of QCD. On the other hand, one expects eventually the
DIGA picture to become reliable for asymptotically high temperatures, even
though one does not know \emph{a priori} exactly when this happens. Therefore, as the temperature is increased, one should see a change in $\theta$-dependence when going from the low-temperature to the high-temperature regime, where the $T$-, $N$- and $m$-dependence of
$\chi$ and $b_{2n}$ are rather different than those predicted by
$\chi$PT and large-$N$ arguments. In the pure Yang--Mills case, the theory
undergoes a deconfinement phase transition, see also Sec.\ref{sec:T_theta_phase_diagram}, so it is natural to expect that the change of $\theta$-dependence should also take place at the transition. This is even more natural in the
large-$N$ limit, since in that case $\chi$ vanishes according to DIGA, hence
the change of $\theta$-dependence must be non-analytic, thus associated to a
phase transition. In the full QCD case, the phase transition is replaced by a
rapid chiral-restoration crossover above which the $\SU(2)_{\subA}$ symmetry is
approximately restored, so that this qualitative picture should not drastically
change. These scenarios are exactly those that have been found to be realized from the
lattice.

Let us start from the pure-glue case. Initial indications of the strong
suppression of $\chi$ above the deconfinement transition were found for both
SU(2)~\cite{Teper:1985gi,Teper:1985ek,Alles:1997qe} and
SU(3)~\cite{Alles:1996nm} gauge theories, while below $T_c$ the susceptibility
was found to be almost independent of $T$ up to temperatures very close to
$T_c$. By increasing $N$, the drop above $T_c$ was found to become sharper and
sharper, and the suppression stronger and stronger, in qualitative agreement
with DIGA, predicting a suppression $\chi(T)\sim T^{4-11N/3}$. This was also
quantitatively established in more recent times by directly
computing~\footnote{Since the deconfinement transition is discontinuous
for $N>2$, we have to distiguish the values of $\chi(T)$ obtained when
approaching $T_c$ from above $\chi_{\scriptscriptstyle{d}} = \chi(T_c^+)$ and from below $\chi_{\scriptscriptstyle{c}} =\chi(T_c^-)$.}
$\chi_{\scriptscriptstyle{d}}=\chi(T_c^+)$~\cite{Borsanyi:2022fub,Bonanno:2023hhp},
which is found to be exponentially suppressed as a function of
$N$~\cite{Bonanno:2023hhp}. This is suggestive of a picture where the DIGA
regime is entered sooner and sooner after the transition as $N$ is increased,
eventually leading to $\chi(T\ge T_c^{+})=0$ and $\chi(T\le T_c^{-})=\chi(T=0)$
for $N=\infty$. Further studies of $\chi(T)$ for $N=3$ at much higher
temperatures, up to $\sim10 T_c$~\cite{Berkowitz:2015aua, Borsanyi:2015cka,
Frison:2016vuc, Kitano:2015fla,Borsanyi:2021gqg}, reported a strong power-law
suppression of the topological susceptibility. The observed suppression
is consistent with that predicted by DIGA, $\chi \sim A T^{-7}$, although the
prefactor $A$ was found to be about an order of magnitude larger
than the DIGA one. This behavior seems to favor a scenario in which the
diluteness hypothesis starts to be valid much earlier (i.e., at smaller
temperature) than the 1-loop approximation employed for the DIGA evaluation of
the 1-instanton partition function.

At this stage, it is important to make the following clarification. A
suppression of $\chi$, although a necessary condition, is actually not
sufficient alone to support the validity of DIGA. As a matter of fact, there
are cases where these two facts do not coexist. One example is offered by $2d$
$\CP^{N-1}$ models: in this case, $\chi \sim 1/N \to 0$~\cite{DAdda:1978vbw,
Luscher:1978rn, Witten:1978bc, Campostrini:1991kv, Campostrini:1991tw} in the
large-$N$ limit, but $b_{2n}\sim 1/N^{2n}$~\cite{DelDebbio:2006yuf,
Rossi:2016uce, Bonati:2016tvi, Sugeno:2025exv}, in striking disagreement with
DIGA. Another example is offered by the $\chi$PT predictions reviewed earlier
in Sec.~\ref{sec:chiPT}. Assuming a non-zero up/down quark mass ratio $z$,
$\chi \sim m \to 0$ in the chiral limit, but the $b_{2n}$ coefficients remain
finite and different from the DIGA ones. $\chi$PT predicts a DIGA behavior only
in the limit $z\to 0$ (massless up quark limit with the down quark remaining
massive), as it can be seen by Taylor-expanding Eq.~\eqref{eq:mpi_chiPT} for
small $z$. For this reason, the actual validity of DIGA must also be checked by
going beyond the leading-order term.  In this respect, the pure-glue results of
Ref.~\cite{Bonati:2013tt} represented an important cross check, as in that
study it was shown that $b_2$ and $b_4$ indeed approach the DIGA values
soon after $T_c$, and at a faster rate as
$N$ is increased. Note also that by studying $b_{2n}$ we are directly
testing the diluteness hypothesis, irrespective of the further ingredients
entering the approximate evaluation of the $1-$instanton partition function, as
noted in Sec.~\ref{subsect:DIGA} This behavior was later confirmed by other
independent studies~\cite{Borsanyi:2015cka, Vig:2021oyt, Kovacs:2023vzi,
Yamada:2024vsk}, and was also extended to other theories characterized by a
deconfining transition, such as the Yang--Mills theory with the gauge group
G$_2$~\cite{Bonati:2015uga}. These results thus prove the existence of a
general deep connection between $\theta$-dependence and the confining
properties of the theory.

Let us now move to the full QCD case, where the determination of
$\theta$-dependence and the clarification of the regime of validity of DIGA are
of extreme relevance for axion cosmology~\cite{Preskill:1982cy,Abbott:1982af,
Dine:1982ah, Wantz:2009it, DiLuzio:2020wdo}. As earlier anticipated,
introducing dynamical quarks does not change significantly the overall picture
found in the pure-glue case. Also in full QCD, indeed, several independent
studies have shown that $\chi$ rapidly drops above the chiral
crossover~\cite{Alles:2000cg, Bonati:2015vqz,Borsanyi:2016ksw,
Petreczky:2016vrs, Bonati:2018blm,
Burger:2017xkz,Athenodorou:2022aay,Chen:2022fid, Kotov:2025ilm}.\footnote{Such
drop was also shown to take place around other transitions present in the QCD
phase diagram, such as in the presence of magnetic
fields~\cite{Brandt:2024gso}, or at finite baryon density for two-color
QCD~\cite{Alles:2006ea,Hands:2011hd,Iida:2019rah,Lombardo:2021jpn,Kawaguchi:2023olk,Iida:2024irv,Itou:2025vcy},
with possible exceptions at low temperatures and large values of the baryon
chemical potential~\cite{Iida:2019rah,Iida:2024irv,Itou:2025vcy}.} Also the
quartic coefficient $b_2$ exhibits a fast approach to the DIGA value above the
chiral crossover, although a bit slower than in pure
Yang--Mills~\cite{Bonati:2015vqz,Kotov:2025ilm}.

However, there is an important difference with respect to the pure Yang--Mills
case. Although most of the available results for $\chi(T)$ agree on the fact
that $\chi$ seems to be power-law suppressed with an exponent compatible with
the DIGA (at least for temperatures above a couple of times the crossover
temperature), it should be stressed that a clear consensus on the
actual values of $\chi$ in the high temperature phase of QCD has not been
achieved yet. This reflects the intrinsic challenges of determining $\chi$ from lattice
calculations with light dynamical fermions, challenges that become even more
pronounced at higher temperatures.

First of all, finite cut-off effects related to the explicit breaking of chiral
symmetry at finite lattice spacing are very large, and can potentially lead to
misleading results (as it happened, for example, in the early study
Ref.~\cite{Bonati:2015vqz}). Secondly, the suppression of $\chi$ at large $T$
implies that one must sample a topological charge distribution with a tiny
variance: $\langle Q^2 \rangle = V \chi \ll 1$: sampling $Q$ in these
conditions is very hard because configurations with non-vanishing $Q$ have a very small
 probability of being found. Extremely long Monte Carlo simulations are thus needed to achieve a reasonable statistical
accuracy on $\chi$. In this respect, the multicanonical approach has been shown
to be an effective approach to mitigate this issue, see~\cite{Bonati:2017woi,
Bonati:2018blm, Jahn:2018dke, Athenodorou:2022aay}. Finally, in the
high-temperature phase, cut-off effects are controlled by $aT = 1/N_\tau$, with
$N_\tau$ the number of lattice temporal points. 
To have these cut-off effects under control 
$N_\tau \gtrsim\mathcal{O}(10)$ is typically needed, which in turn requires very fine lattice spacings to keep $T$ large. This exacerbates the
topological freezing problem, and explains why most studies are limited to
$T\lesssim 4 T_c$, while axion phenomenology would require at least
temperatures up to $T\sim 10 T_c$. Algorithmic development for topological
freezing is currently a very active research field, and several different
approaches are under scrutiny, either in lower-dimensional models or in $4d$
gauge theories and in QCD, see, e.g.,
Refs.~\cite{Luscher:2011kk, LSD:2014yyp, Mages:2015scv, Bietenholz:2015rsa,
Laio:2015era, Bonati:2017woi, Hasenbusch:2017unr, Luscher:2017cjh,
Bonanno:2020hht, Florio:2019nte, Funcke:2019zna, Kanwar:2020xzo,
Nicoli:2020njz, Albandea:2021lvl, Cossu:2021bgn, Borsanyi:2021gqg,
Fritzsch:2021klm, Abbott:2023thq, Eichhorn:2023uge, Howarth:2023bwk,
Albandea:2024fui, Bonanno:2024udh, Abe:2024fpt, Bonanno:2024zyn,
Bonanno:2025pdp, Katz:2025ycr}.

\subsection{The neutron EDM from lattice QCD}

%riassunto determinazioni da reticolo

\begin{figure}[!t]
\centering
\includegraphics[scale=0.55]{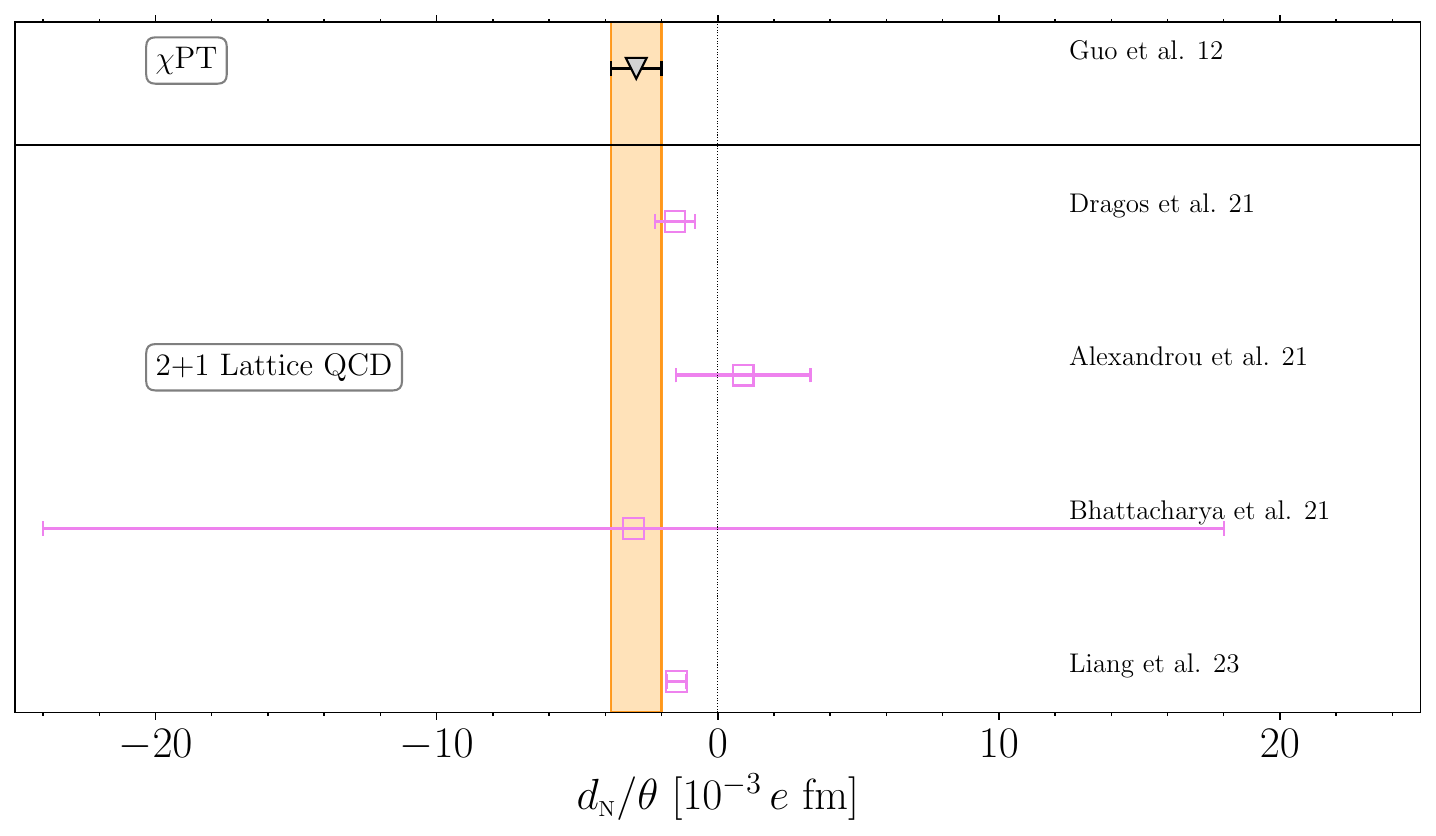}
\caption{Summary of neutron EDM determinations from $2+1$ lattice QCD~\cite{Dragos:2019oxn,Alexandrou:2020mds,Bhattacharya:2021lol,Liang:2023jfj} (square points), compared with the $\chi$PT result~\cite{Guo:2012vf} (triangle point). The EDM is reported in units of $10^{-3} \cdot \theta~e~\mathrm{fm}$, with $e$ the elementary electric charge.}
\label{fig:nEDM_comp}
\end{figure}

In recent years, Lattice QCD calculations have 
reached the accuracy needed to provide a first principle evaluation of the 
neutron EDM $d_{\subN}$ as function of $\theta$.
On the lattice, $d_{\subN}$ is computed from a specific CP-odd
nucleon electromagnetic form factor called
$F_3$~\cite{Shintani:2005xg,Liu:2008gr,Abramczyk:2017oxr}, which can be
extracted from the CP-odd component of the nucleon matrix element of the
electromagnetic quark current
$\bra{\mathrm{N}(P)}J_{\EM}\ket{\mathrm{N}(P^\prime)}$. Since the form factor
$F_3$ depends on the exchanged momentum $\Delta P^2 = -(P^\prime - P)^2$
between the initial and final nucleon states in the matrix element, one has to
take the limit $\Delta P\to 0$ to actually obtain the static EDM. In formulas, this is
written as~\cite{Pospelov:2005pr}:
\beq
d_{\subN}(\theta) = \lim_{\Delta P^2 \, \to \, 0} \frac{F_3(\Delta P^2,\theta)}{2M_{\subN}},
\eeq
with $M_{\subN}$ the nucleon mass, extracted on the lattice from the large-time
decay of the nucleon-nucleon two-point correlator. 
 Being $d_{\subN}$ CP-odd, one
expects on general grounds $d_{\subN}(\theta) = C \theta\left[1 +
\mathcal{O}(\theta)\right]$. Therefore it is sufficient to compute just the coefficient $C$, which is equal to $C = d_{\subN}(\theta)/\theta$ at leading order in $\theta$. The value of $C$ can be extracted from the large Euclidean
time separation of an appropriate correlator, built via the insertion of the
topological charge in the EDM correlation function:
\beq
\braket{O_{\EDM}(\tau_{\scriptscriptstyle{\rm sep}})}_\theta = \braket{O_{\EDM}(\tau_{\scriptscriptstyle{\rm sep}})}_{\theta \, = \, 0} + \ii \theta \braket{O_{\EDM}(\tau_{\scriptscriptstyle{\rm sep}}) Q}_{\theta \, = \, 0} + \mathcal{O}(\theta^2) = \ii \theta \braket{O_{\EDM}(\tau_{\scriptscriptstyle{\rm sep}}) Q}_{\theta \, = \, 0} + \mathcal{O}(\theta^2),
\eeq
where $O_{\EDM}(\tau_{\scriptscriptstyle{\rm sep}})$ is the interpolating
operator for the nucleon EDM. Schematically:
$O_{\EDM}(\tau_{\scriptscriptstyle{\rm
sep}})=J_{\subN}(P,\tau_{\scriptscriptstyle{\rm
f}})J_{\EM}\bar{J}_{\subN}(P^\prime,\tau_{\scriptscriptstyle{\rm i}})$, with
$J_{\EM}$ the quark electromagnetic current, $J_{\subN}(P,\tau)$ the Euclidean
time and momentum dependent nucleon interpolating operator, and
$\tau_{\scriptscriptstyle{\rm sep}}=\tau_{\scriptscriptstyle{\rm
f}}-\tau_{\scriptscriptstyle{\rm i}}$ the time separation between the two
nucleon interpolating operators. The insertion of the topological charge makes
the extraction of $d_{\subN}(\theta)/\theta$ from
$\braket{O_{\EDM}(\tau_{\scriptscriptstyle{\rm sep}})Q}_{\theta \, = \, 0}$
very noisy, with a bad scaling of the signal-to-noise ratio when the volume is
increased (much like what happens with the $b_{2n}$ coefficients).
On the other hand, the use of sufficiently large volumes is mandatory for this
calculation, as excited state contamination in the nucleon channel can be quite large~\cite{Bhattacharya:2021lol}, and thus needs to be damped by
evaluating $\braket{O_{\EDM}(\tau_{\scriptscriptstyle{\rm sep}}) Q}_{\theta
\, = \, 0}$ at very large Euclidean time separations.

A summary of lattice QCD determinations of the neutron EDM is reported in
Fig.~\ref{fig:nEDM_comp}. As it can be seen, they are all in good agreement
with the $\chi$PT prediction. Currently, the most accurate determination ~\cite{Liang:2023jfj} still lacks a continuum extrapolation, but improvements are
expected in the near future. For a recent review on the topic and more
technical details on the extraction of $d_{\subN}$ from the lattice, we refer
the reader to the recent review~\cite{Liu:2024kqy}.

\subsection{The sphaleron rate from lattice QCD}

The sphaleron rate as defined in Sec.~\ref{sec:intro_sphaleron} is the
space-time integral of a two-point function in \emph{real-time}, which is a
quantity not directly accessible in lattice investigations based on the
Euclidean formulation of QCD. There is however a standard way to
relate $\Gamma_\S$ to Euclidean quantities, which also works for other
kinetic coefficients. Let us introduce the spectral energy density
$\rho(\omega)$ of the Euclidean time correlator of the topological charge
density at zero spatial momentum $G_{\subE}(\tau)$:
\beq
\label{eq:eucl_spec_dens_def}
G_{\subE}(\tau) = \int \dd^3 x \braket{q(\vec{x},\tau)q(\vec{0},0)}_{\subE} = -\int_0^{\infty} \frac{\dd\omega}{\pi} \rho(\omega) \frac{\cosh\left(\frac{\omega}{2T}-\omega \tau\right)}{\sinh\left(\frac{\omega}{2T}\right)} \, ,
\eeq
where the notation $\braket{O}_{\subE}$ is used to recall that
this is an expectation value computed in Euclidean space-time with a
compactified time direction with length equal to $1/T$. Performing an analytic continuation, one can eventually relate $\Gamma_{\S}$ to the zero-energy limit $\omega\to 0$ of the slope of the Euclidean spectral density~\cite{Meyer:2011gj,Lowdon:2022keu}:
\beq\label{eq:kubo_formula}
\Gamma_{\S} = 2T \lim_{\omega\,\to\,0} \frac{\rho(\omega)}{\omega} \, .
\eeq
This formula, taken at face value, does not make the task of computing
$\Gamma_{\S}$ any easier: on the lattice one can only compute $G_{\subE}(\tau)$
for a discrete set of points and with statistical uncertainties, and obtaining
$\rho(\omega)$ from $G_{\subE}(\tau)$ requires to solve an
ill-conditioned inverse problem, where the propagation of errors from
$G_{\subE}(\tau)$ to $\rho(\omega)$ is out of control unless a 
regularization prescription is adopted.

\begin{figure}[!t]
\centering
\includegraphics[scale=0.55]{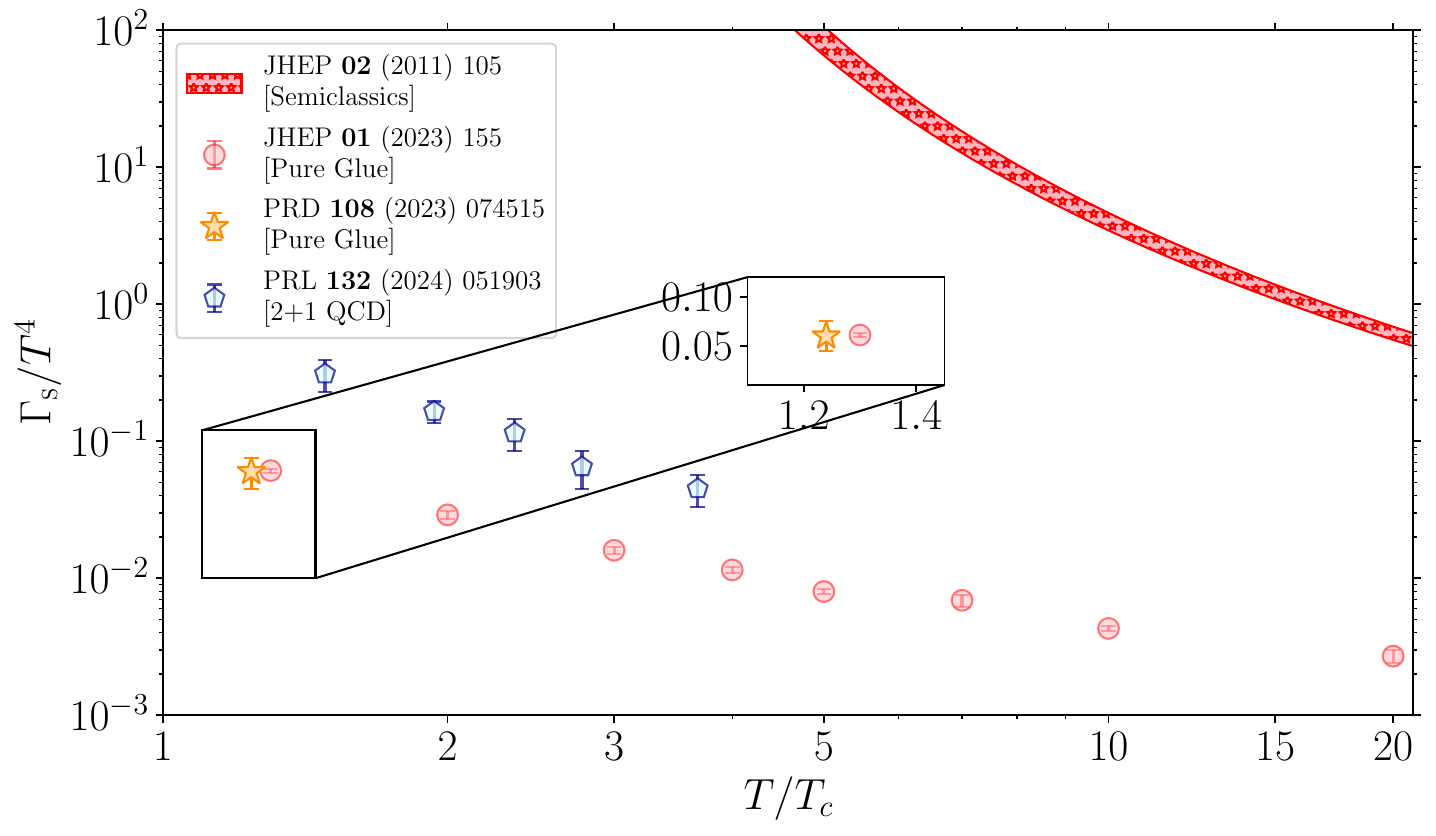}
\caption{Comparison among the non-perturbative pure-glue and 2+1 QCD determinations of the sphaleron rate of~\cite{Bonanno:2023ljc,Bonanno:2023thi} with the new method~\cite{Bonanno:2023ljc} that addresses the resolution of the inverse problem, the non-perturbative pure-glue determination of~\cite{BarrosoMancha:2022mbj} that avoided the resolution of the inverse problem, and the semiclassical estimate of~\cite{Moore:2010jd}.}
\label{fig:sphal_rate}
\end{figure}

In recent years, there has been significant progress in lattice methods to
solve inverse problems~\cite{Astrakhantsev:2018oue, Hansen:2019idp,
Boito:2022njs, Rothkopf:2022fyo, Altenkort:2023oms, Bruno:2024fqc,
DelDebbio:2024lwm}.  In particular, a novel proposal by Hansen, Lupo and
Tantalo (HLT)~\cite{Hansen:2019idp} has been particularly successful, and has been used in several papers to solve inverse problems appearing in vastly different physical
contexts, ranging from Beyond Standard Model particle spectroscopy to flavor
physics~\cite{ExtendedTwistedMassCollaborationETMC:2022sta, DelDebbio:2022qgu,
Frezzotti:2023nun, Evangelista:2023fmt, ExtendedTwistedMass:2024myu,
Bennett:2024cqv, DeSantis:2025qbb, DeSantis:2025yfm, Frezzotti:2025hif,
TELOS:2025ash}. This method is a refinement of a rather well-established strategy to
find approximate solutions of inverse problems proposed long ago by Backus and
Gilbert (BG)~\cite{BackusGilbert1968:aaa}, which regularizes the inverse
problem by means of a smearing procedure. In the limit in which the width of
the kernel used for the smearing tends to zero, one
recovers the exact solution. The main shortcoming of BG is that the smearing
width is an \emph{output} of the method and cannot be controlled, thus
preventing an accurate assessment of the systematic error associated to the
smearing regularization. Instead, HLT devised a modification of BG that allows
to choose the smearing width as an \emph{input} of the method, allowing to
actually perform a controlled zero-smearing-width numerical extrapolation. This
extrapolation from the HLT method, accompanied by the continuum limit and by the
zero-smoothing-limit to remove any unphysical effect in $G_{\subE}(\tau)$ (much
like what has been done for $\chi^\prime$), are the fundamental ingredients of
the novel proposal put forward in~\cite{Bonanno:2023ljc} to compute the
sphaleron rate from the lattice. This new method has allowed the first
non-perturbative determination of $\Gamma_\S$ in 2+1 QCD with physical quarks
at high temperatures~\cite{Bonanno:2023thi}.

The full QCD data of~\cite{Bonanno:2023thi} are shown in
Fig.~\ref{fig:sphal_rate}, where they are compared with the pure-glue results
of~\cite{BarrosoMancha:2022mbj}. As it can be seen, the inclusion of the
dynamical quarks has the effect of increasing $\Gamma_{\S}$. Since the pure-glue and the full QCD theory are characterized by different temperature scales that separate the low-$T$ and the high-$T$ regimes, in this plot we show the results for the sphaleron rate as a function of $T/T_c$, where $T_c\simeq 155$ MeV was used in full QCD (chiral crossover temperature), while $T_c\simeq 287$ MeV was used in the pure-glue theory (critical
deconfinement temperature), see~\cite{Bonanno:2023thi} for more
details. On the other hand, the semiclassical prediction
of~\cite{Moore:2010jd},
\beq
\frac{\Gamma_{\S}}{T^4} \propto \alpha_s^5, \qquad \qquad \alpha_s = \frac{g^2}{4\pi},
\eeq
is unreliable in the explored range of temperatures, where non-perturbative
effects are still large (see also~\cite{Guin:2026kbp}). Given the phenomenological relevance of the sphaleron rate, see Sec.~\ref{sec:intro_sphaleron}, it would be extremely important to clearly
assess the temperature dependence of $\Gamma_\S$, however this requires the
exploration of a large temperature range, see the discussion
in~\cite{Bonanno:2023thi}.

\subsection{Confinement at non-zero \texorpdfstring{$\theta$}{theta} and the \texorpdfstring{$T-\theta$}{T-theta} phase diagram}\label{sec:T_theta_phase_diagram}

A rather interesting and non-trivial question is how a non-zero $\theta$ would
alter the confining properties of the theory.
At $\theta=0$, it is well known that, as the temperature is increased, $\SU(N)$
pure-glue theories undergo a phase transition at a critical temperature $T_c$.
Below $T_c$, Yang--Mills theories exhibit color confinement, and their expectation values are invariant under gauge transformations that are periodic up to an element of the center of the gauge group $\mathbb{Z}_N$ (center symmetry, for short). Above $T_c$ this is no more the case, and the theory deconfines. The Polyakov loop, the holonomy of the gauge field around the compactified time direction, is the order parameter for center symmetry breaking: it averages to zero in the symmetric phase, while it aligns along one of the $N$ center elements of $\mathbb{Z}_N$ in the broken phase.

The deconfinement transition is discontinuous for $N\ge
3$~\cite{Fukugita:1989yw, Fingberg:1992ju, Beinlich:1997ia, Campostrini:1998zd, Lucini:2001ej, Lucini:2002ku, Lucini:2003zr, Lucini:2004yh, Lucini:2005vg, Lucini:2012wq, Borsanyi:2022xml, Lucini:2023irm,Cohen:2023hbq},
thus, at $T=T_c$, one has a phase coexistence and $f_c(T_c)=f_d(T_c)$, with $f_c,f_d$ denoting the free energies in the confined and deconfinement phase. Increasing $\theta$, one expects the transition to remain first-order, however, one can also predict that the temperature at which $f_c=f_d$ will decrease as a function of $\theta$~\cite{Unsal:2012zj,Poppitz:2012nz,Anber:2013sga,Kitano:2017jng,Aitken:2018mbb,Chen:2020syd}.
Indeed, we have seen that $F(T,\theta)=f(T,\theta)-f(T,0)=\frac{1}{2}\chi(T)\theta^2 +
\mathcal{O}(\theta^4)$, and we have discussed that $\chi$ is rapidly suppressed
at high temperatures, above the transition, while $\chi$ is almost constant
below $T_c$. Thus, due to the sharp drop of $\chi$ above $T_c$, the matching
condition $f_c(T_c(\theta),\theta)=f_d(T_c(\theta),\theta)$ will take place for
a lower temperature compared to $\theta=0$. Developing this argument one
arrives at the following expression:
\beq
T_c(\theta) = T_c(0)\left[ 1 - R \theta^2 + \mathcal{O}(\theta^4)\right]\, , \qquad R = \frac{\Delta \chi}{2L} > 0\, , \qquad L = (\epsilon_{\scriptscriptstyle{d}} - \epsilon_{\scriptscriptstyle{c}})\big\vert_{\theta\,=\, 0} \, , \quad \Delta \chi = (\chi_{\scriptscriptstyle{c}} - \chi_{\scriptscriptstyle{d}})\big\vert_{\theta\,=\, 0} \, .
\eeq
Here, $L$ denotes the latent heat of the $\theta=0$ transition (i.e., the jump
of the internal energy density), while $\Delta \chi$ represents the
jump of the topological susceptibility at the $\theta=0$ transition. The
decrease of $T_c(\theta)$ was investigated using simulations performed at imaginary $\theta$ values~\cite{DElia:2012pvq,DElia:2013uaf,Otake:2022bcq,Borsanyi:2022xml, Bonanno:2023hhp}. 
The large-$N$ behavior of the slope $R$ was shown~\cite{Bonanno:2023hhp}
to be consistent with the scaling form:
\beq
R = \frac{\bar{R}}{N^2} + \mathcal{O}\left(\frac{1}{N^4}\right) \, ,
\eeq
as expected on general theoretical grounds, since at large-$N$ the actual
expansion variable is $\bar{\theta}=\theta/N$, and also from the fact that the
latent heat is proportional to the number of degrees of freedom, hence to
$N^2$, while $\Delta\chi$ approaches a constant value in the large-$N$ limit. The $\theta$-dependence of the critical temperature  has also been investigated in the $\SU(2)$ case, where the transition is second-order~\cite{Yamada:2024pjy}.  In that case, the
relation between $R$ and $\Delta \chi/(2L)$ does not hold of course; still, it
is found that $R>0$ even in this case~\cite{Kitano:2020mfk,Kitano:2021jho,Yamada:2024vsk}. This is once again a confirmation of the fact that the $\theta$-dependence of the $\SU(2)$
theory is qualitatively analogous to that of the $N\ge 3$
theories~\cite{Kitano:2020mfk,Kitano:2021jho,Yamada:2024vsk}.

Interestingly, a similar $\mathcal{O}(\theta^2)$ decrease has also been
found in other scales related to confinement in Yang--Mills
theories, namely, the mass gap $m_{\subG}$ and the string tension $\sigma$. In
those cases, lattice simulations have shown
that~\cite{DelDebbio:2006yuf,Bonanno:2024ggk}:
\beq
m_{\subG}(\theta) = m_{\subG}(0)\left[ 1 + m_2 \theta^2 + \mathcal{O}(\theta^4)\right] \, , \qquad \qquad \sigma(\theta) = \sigma(0)\left[ 1 + s_2 \theta^2 + \mathcal{O}(\theta^4)\right] \, ,
\eeq
where it has been found that~\cite{Bonanno:2024ggk}
%\
\beq
m_2 = \frac{\bar{m}_2}{N^2} + \mathcal{O}\left(\frac{1}{N^4}\right), \qquad \qquad
s_2 = \frac{\bar{s}_2}{N^2} + \mathcal{O}\left(\frac{1}{N^4}\right), \qquad \qquad  \bar{m}_2, \bar{s}_2 < 0.
\eeq

\section{Conclusions}\label{sec:conclu}

In this chapter, we have provided a pedagogical introduction to
$\theta$-dependence in QCD. We discussed its salient theoretical features and
phenomenological implications, with particular focus on hadron phenomenology
and axion cosmology. Then, we moved to the discussion of available analytical
and numerical results. We reviewed analytic predictions obtained from chiral
effective theories, large-$N$ arguments and semiclassical methods. Concerning
lattice results, we conducted a summary of state-of-the-art results, with a selection of the most recent ones being collected in Tab.~\ref{tab:summary_main_results}.

Despite significant progress in the last decade, there are still many questions
that to this day remain unanswered, and that deserve to be further investigated
in the near future. The most important short-term goal is perhaps the one related to reaching a quantitative consensus on the actual value of the topological susceptibility in the high-temperature phase, and clarify the origin of the existing discrepancies among results
obtained by different groups. As discussed in Sec.~\ref{sec:HT}, topological
freezing is one of the major obstacle towards this objective, and algorithmic
development will be key to advance this research direction; at the same
time, novel lattice discretizations could make the need to reach small lattice
spacings less stringent. As already pointed out, this is at present a very
active field of research, and several novel approaches to mitigate freezing are
under development, see, e.g., Refs.~\cite{Luscher:2011kk,
LSD:2014yyp,Mages:2015scv, Bietenholz:2015rsa, Laio:2015era, Bonati:2017woi,
Hasenbusch:2017unr, Luscher:2017cjh, Bonanno:2020hht,
Florio:2019nte,Funcke:2019zna, Kanwar:2020xzo, Nicoli:2020njz,
Albandea:2021lvl, Cossu:2021bgn, Borsanyi:2021gqg, Fritzsch:2021klm,
Abbott:2023thq, Eichhorn:2023uge, Howarth:2023bwk, Albandea:2024fui,
Bonanno:2024udh, Abe:2024fpt, Bonanno:2024zyn, Bonanno:2025pdp}. Another very
important short-term research direction that deserves to be pursued is the
study of real-time dynamics of topological fluctuations in QCD. So far, only
results for the sphaleron rate are available~\cite{Kotov:2018vyl,
Altenkort:2020axj, BarrosoMancha:2022mbj, Bonanno:2023ljc,
Bonanno:2023thi,Guin:2026kbp}, with only one study addressing the full QCD
calculation. Extension to non-zero energies and higher temperatures is surely a
future outlook that ought to be pursued.

Finally, there are also a few theoretical research directions with strong
tights to QCD topology that are of current interest, and that are being carried
on. One regards the clarification of the connection between QCD topological
features and the infrared behavior of the Dirac spectrum in the
high-temperature phase of QCD. This is particularly important to clarify the
debated effective restoration of the anomalous $\U(1)_{\subA}$ symmetry above
the chiral crossover~\cite{Edwards:1999zm, Dick:2015twa, Aoki:2020noz,
Ding:2020xlj, Aoki:2021qws, Kaczmarek:2021ser, Kaczmarek:2023bxb,
Kovacs:2023vzi, Azcoiti:2023xvu, Giordano:2024jnc, Alexandru:2024tel,
Fodor:2025yuj,Aarts:2026kpq}. Another one regards instead the behavior of
$\theta$-dependence for large $\theta$ angles, i.e., close to $\theta=\pi$.
Indeed, the results reviewed here all refer to the behavior for small $\theta$
angles, i.e., around $\theta=0$, which is the regime realized
by real-world QCD. However, theoretically it is very interesting to clarify how
$\theta$-dependence changes close to $\theta\sim\pi$, especially because novel
features of the theory could emerge. For example, we have discussed the
first-order phase transition that is expected to take place in the large-$N$
limit for $\theta=\pi$. Clearly, direct investigations of the large-$\theta$
case are hindered by the sign problem, and methods based on analytic
continuation are expected to experience a rapidly degrading signal-to-noise
ratio when pushing them towards larger and larger values of $\theta$.
Currently, there are ongoing studies testing novel numerical techniques to
overcome these issues both in $\SU(2)$ gauge theories~\cite{Kitano:2017jng,
Kitano:2020mfk, Hirasawa:2024fjt} and in lower-dimensional
models~\cite{Alles:2007br, Alles:2014tta, Sulejmanpasic:2020lyq,Sugeno:2025exv,
Matsumoto:2025fjb}.

%%% Final Table %%%

\begin{table}[!t]
\begin{center}
\begin{tabular}{|c|}
\hline
\textbf{$\theta$-dependence of the vacuum energy $E_{\subG}(\theta)$ in $\SU(N)$ pure-gauge theories}\\
\\[-1em]
\hline
\\[-1em]
$E_{\subG}(\theta)=\frac{1}{2}\chi\theta^2\left[1 + b_2 \theta^2 + b_4\theta^4 + \mathcal{O}(\theta^6)\right]$  $\qquad \qquad$ $\chi \to $ \textbf{Topological Susceptibility}\\
\hline\\[-1em]
\\[-1em]
$
\begin{aligned}
&\dfrac{\chi}{\sigma^2} = 0.02941(21)~\text{\cite{Bonanno:2025eeb}} \qquad\qquad &(8t_0)^2 \chi = 0.04499(83)~\text{\cite{Ce:2015qha}} \qquad\qquad &\chi^{1/4} = 179.31(75)~\mathrm{MeV}~\text{\cite{Bonanno:2025eeb}} \qquad\qquad &(N=3) \,\,\,\,\\
\\[-1em]
&\dfrac{\chi}{\sigma^2} = 0.02088(39)~\text{\cite{Bonanno:2025eeb}} \qquad\qquad &(8t_0)^2 \chi = 0.04269(45)~\text{\cite{Ce:2016awn}} \qquad\qquad &\chi^{1/4}=180.94(84)~\mathrm{MeV}~\text{\cite{Bonanno:2025eeb}} \qquad\qquad &(N\to\infty)\\
\end{aligned}
$
\\
\\
$\dfrac{\chi}{\sigma^2} = 0.02088(39) + 0.044(12)\dfrac{1}{N^2} + 0.293(83) \dfrac{1}{N^4} + \mathcal{O}\left(\dfrac{1}{N^6}\right)$~\text{\cite{Bonanno:2025eeb}}\\
\\[-1em]
\\
$b_2= -0.0216(15)$~\text{\cite{Bonati:2015sqt}} $\quad$ $(N=3)$ $\qquad\qquad\qquad\qquad$ $b_2 = -0.1931(98) \dfrac{1}{N^2}$~\text{\cite{Bonanno:2020hht}} $\quad$ $(N\ge 3)$\\
\\[-1em]
\\
$\dfrac{\chi}{\sigma^2} = 0.05190(61)~\text{\cite{Athenodorou:2021qvs}}$ $\qquad\qquad$ $b_2 = -0.0315(52)$~\text{\cite{Bonanno:2018xtd}} $\qquad\qquad$ $b_4=(5.8 \pm 1.9) \times 10^{-4}$~\text{\cite{Bonanno:2018xtd}} $\qquad\qquad(N=2)$\\[-1em]
\multicolumn{1}{|c|}{}\\\hline
\hline
\textbf{$\theta$-dependence of non-topological quantities in $\SU(N)$ pure-gauge theories}\\
\hline\\[-1em]
$
\begin{aligned}
&\textbf{Mass gap} \qquad\qquad &\dfrac{m_{\subG}(\theta)}{m_{\subG}(0)} &= 1 -0.075(20)\dfrac{\theta^2}{N^2}~\text{\cite{Bonanno:2024ggk}} \qquad\qquad &(N\ge3)\\
\\[-1em]
\\[-1em]
&\textbf{String tension} \qquad\qquad &\dfrac{\sigma(\theta)}{\sigma(0)} &=  1 - 0.23(1)\dfrac{\theta^2}{N^2}~\text{\cite{Bonanno:2024ggk}} \qquad\qquad &(N\ge3)\\
\\[-1em]
\\[-1em]
&\textbf{Deconfinement temperature} \qquad\qquad &\dfrac{T_c(\theta)}{T_c(0)} &= 1 - 0.159(4)\dfrac{\theta^2}{N^2}~\text{\cite{DElia:2012pvq,Bonanno:2023hhp}} \qquad\qquad &(N\ge3)\\
\end{aligned}
$
\\[-1em]
\multicolumn{1}{|c|}{}\\\hline
\hline
\textbf{$\theta$-dependence of the vacuum energy in QCD with physical quark masses}\\
\hline
\\[-1em]
$
\begin{aligned}
&\chi^{1/4} = (75.5 \pm 3.4)~\mathrm{MeV}~\text{\cite{Bruno:2014ova}} \qquad\qquad &(N_{\subf}=2) \quad\qquad &\\
\\[-1em]
&\chi^{1/4} = (79.6 \pm 3.5)~\mathrm{MeV}~\text{\cite{Bonanno:2024zyn}} \qquad\qquad &(N_{\subf}=2+1) \,\,\,\quad &\\
\\[-1em]
&\chi^{1/4} = (78.1 \pm 2.2)~\mathrm{MeV}~\text{\cite{Borsanyi:2016ksw}} \qquad\qquad &(N_{\subf}=2+1+1)  &
\end{aligned}
$
\\[-1em]
\multicolumn{1}{|c|}{}\\\hline
\hline
\textbf{Neutron Electric Dipole Moment in QCD with physical quark masses}\\
\hline
\\[-1em]
\\[-1em]
$d_{\subN}(\theta) = (-1.48 \pm 0.34) \times 10^{-3}~\theta\,e\,\mathrm{fm}$~\text{\cite{Liang:2023jfj}} $\qquad\qquad(N_{\subf}=2+1)$\\[-1em]
\multicolumn{1}{|c|}{}\\
\hline
\end{tabular}
\end{center}
\caption{Summary of the main lattice QCD results reviewed in this chapter.}
\label{tab:summary_main_results}
\end{table}

\begin{ack}[Acknowledgments]
This work is partially supported by the project ``Non-perturbative aspects of fundamental interactions, in the Standard Model and beyond'' funded by MUR, Progetti di Ricerca di Rilevante Interesse Nazionale (PRIN), Bando 2022, grant 2022TJFCYB (CUP I53D23001440006).
\end{ack}

%%%%%%%%%%%%%%%%%%%%%%%%%%%%%%%%%%%%%%%%%%%%
\seealso{
CP violation and Electric Dipole Moments (theory);
The neutron EDM;
Axions and axion-like particles;
Chiral and conformal anomaly in high energy QCD;
Low-energy QCD Effective Models;
Topological Configurations and Color Confinement;
Chiral Magnetic Effect and Related Anomalous Transport Phenomena;
QCD-Like Theories with Different Color Numbers;
Effective Restoration of the Axial U(1) Symmetry;
Lattice QCD at Finite Temperature and Density;
}

%%%%%%%%%%%%%%%%%%%%%%%%%%%%%%%%%%%%%%%%%

\end{document}